\documentclass{sn-jnl}
\usepackage{aas_macros}
\usepackage{setspace}

\usepackage{url}
\usepackage{hyperref}
\usepackage{color}

\renewcommand{\apss}{Astrophys. Space Sci.}

\newcommand{\degr}{\hbox{$^{\circ}$}}

\newcommand{\farcmin}{.\!\!$^{\prime}$}
\newcommand\fdg{\mbox{$.\!\!^\circ$}}%
\newcommand\arcmin{\mbox{$^{\prime}$}}%
\newcommand{\unitvect}[1]{\mbox{$\hat{#1}$}}
\newcommand{\unit}[1]{\hat{\mathbf{#1}}}

\renewcommand{\figurename}{{\bf Figure}}
\renewcommand{\tablename}{{\bf Table}}

\begin{document}

\title{Determination of X-ray pulsar geometry with IXPE polarimetry}

\author*[1]{Victor Doroshenko}\email{doroshv@astro.uni-tuebingen.de}
\equalcont{These authors contributed equally to this work.}
\author*[2,3]{Juri Poutanen}\email{juri.poutanen@utu.fi}
\equalcont{These authors contributed equally to this work.}
\author[2,3]{Sergey S. Tsygankov}
\author[1]{Valery F. Suleimanov}
\author[4]{Matteo Bachetti}
\author[5]{Ilaria Caiazzo}
\author[6]{Enrico Costa}
\author[6]{Alessandro Di Marco}
\author[7]{Jeremy Heyl}
\author[6]{Fabio La Monaca}
\author[6]{Fabio Muleri}
\author[8,9]{Alexander A. Mushtukov}
\author[10]{George G. Pavlov}
\author[11]{Brian D. Ramsey}
\author[6]{John Rankin}
\author[1]{Andrea Santangelo}
\author[6]{Paolo Soffitta}
\author[1]{R\"udiger Staubert}
\author[11]{Martin C. Weisskopf}
\author[12]{Silvia Zane}
\author[13]{Iv\'an Agudo}
\author[14,15]{Lucio A. Antonelli}
\author[16,17]{Luca Baldini}
\author[11]{Wayne H. Baumgartner}
\author[16]{Ronaldo Bellazzini}
\author[18]{Stefano Bianchi}
\author[11]{Stephen D. Bongiorno}
\author[19,20]{Raffaella Bonino}
\author[16]{Alessandro Brez}
\author[21,22,23]{Niccol\`o Bucciantini}
\author[6]{Fiamma Capitanio}
\author[16]{Simone Castellano}
\author[24]{Elisabetta Cavazzuti}
\author[25,15]{Stefano Ciprini}
\author[6]{Alessandra De Rosa}
\author[6]{Ettore Del Monte}
\author[24]{Laura Di Gesu}
\author[26]{Niccol\`o Di Lalla}
\author[24]{Immacolata Donnarumma}
\author[27]{Michal Dov\v{c}iak}
\author[11]{Steven R. Ehlert}
\author[28]{Teruaki Enoto}
\author[6]{Yuri Evangelista}
\author[6]{Sergio Fabiani}
\author[6]{Riccardo Ferrazzoli}
\author[29]{Javier A. Garcia}
\author[30]{Shuichi Gunji}
\author[31]{Kiyoshi Hayashida$\ddag$}
\author[32]{Wataru Iwakiri}
\author[33,34]{\fnm{Svetlana G.} \sur{Jorstad}}
\author[27]{Vladimir Karas}
\author[28]{Takao Kitaguchi}
\author[11]{Jeffery J. Kolodziejczak}
\author[35]{Henric Krawczynski}
\author[19]{Luca Latronico}
\author[36]{Ioannis Liodakis}
\author[19]{Simone Maldera}
\author[16]{Alberto Manfreda}
\author[37]{Fr\'ed\'eric Marin}
\author[24]{Andrea Marinucci}
\author[33]{Alan P. Marscher}
\author[38]{Herman L. Marshall}
\author[18]{Giorgio Matt}
\author[39]{Ikuyuki Mitsuishi}
\author[40]{Tsunefumi Mizuno}
\author[41]{C.-Y. Ng}
\author[11]{Stephen L. O'Dell}
\author[26]{Nicola Omodei}
\author[19]{Chiara Oppedisano}
\author[14]{Alessandro Papitto}
\author[26]{Abel L. Peirson}
\author[15,14]{Matteo Perri}
\author[16]{Melissa Pesce-Rollins}
\author[4]{Maura Pilia}
\author[4]{Andrea Possenti}
\author[15]{Simonetta Puccetti}
\author[6]{Ajay Ratheesh}
\author[26]{Roger W. Romani}
\author[16]{Carmelo Sgr\`o}
\author[42]{Patrick Slane}
\author[16]{Gloria Spandre}
\author[43,3]{Rashid A. Sunyaev}
\author[28]{Toru Tamagawa}
\author[44]{Fabrizio Tavecchio}
\author[45]{Roberto Taverna}
\author[39]{Yuzuru Tawara}
\author[11]{Allyn F. Tennant}
\author[11]{Nicolas E. Thomas}
\author[46,25,47]{Francesco Tombesi}
\author[4]{Alessio Trois}
\author[45,12]{Roberto Turolla}
\author[48]{Jacco Vink}
\author[12]{Kinwah Wu}
\author[49,6]{Fei Xie}
\affil[1]{Institut f\"ur Astronomie und Astrophysik, Universit\"at T\"ubingen, Sand 1, 72076 T\"ubingen, Germany}
\affil[2]{Department of Physics and Astronomy, 20014 University of Turku, Finland}
\affil[3]{Space Research Institute of the Russian Academy of Sciences, Profsoyuznaya Str. 84/32, Moscow 117997, Russia}
\affil[4]{INAF Osservatorio Astronomico di Cagliari, Via della Scienza 5, 09047 Selargius (CA), Italy}
\affil[5]{TAPIR, Walter Burke Institute for Theoretical Physics, Mail Code 350-17, Caltech, Pasadena, CA 91125, USA}
\affil[6]{INAF Istituto di Astrofisica e Planetologia Spaziali, Via del Fosso del Cavaliere 100, 00133 Roma, Italy}
\affil[7]{University of British Columbia, Vancouver, BC V6T 1Z4, Canada}
\affil[8]{Astrophysics, Department of Physics, University of Oxford, Denys Wilkinson Building, Keble Road, Oxford OX1 3RH, UK}
\affil[9]{Leiden Observatory, Leiden University, NL-2300 RA Leiden, The Netherlands}
\affil[10]{Department of Astronomy and Astrophysics, Pennsylvania State University, University Park, PA 16802, USA}
\affil[11]{NASA Marshall Space Flight Center, Huntsville, AL 35812, USA}
\affil[12]{Mullard Space Science Laboratory, University College London, Holmbury St Mary, Dorking, Surrey RH5 6NT, UK}
\affil[13]{Instituto de Astrof\'isica de Andaluc\'ia—CSIC, Glorieta de la Astronom\'ia s/n, 18008 Granada, Spain}
\affil[14]{INAF Osservatorio Astronomico di Roma, Via Frascati 33, 00078 Monte Porzio Catone (RM), Italy}
\affil[15]{Space Science Data Center, Agenzia Spaziale Italiana, Via del Politecnico snc, 00133 Roma, Italy}
\affil[16]{Istituto Nazionale di Fisica Nucleare, Sezione di Pisa, Largo B. Pontecorvo 3, 56127 Pisa, Italy}
\affil[17]{Dipartimento di Fisica, Universit\`a di Pisa, Largo B. Pontecorvo 3, 56127 Pisa, Italy}
\affil[18]{Dipartimento di Matematica e Fisica, Universit\`a degli Studi Roma Tre, Via della Vasca Navale 84, 00146 Roma, Italy}
\affil[19]{Istituto Nazionale di Fisica Nucleare, Sezione di Torino, Via Pietro Giuria 1, 10125 Torino, Italy}
\affil[20]{Dipartimento di Fisica, Universit\`a degli Studi di Torino, Via Pietro Giuria 1, 10125 Torino, Italy}
\affil[21]{INAF Osservatorio Astrofisico di Arcetri, Largo Enrico Fermi 5, 50125 Firenze, Italy}
\affil[22]{Dipartimento di Fisica e Astronomia, Universit\`a degli Studi di Firenze, Via Sansone 1, 50019 Sesto Fiorentino (FI), Italy}
\affil[23]{Istituto Nazionale di Fisica Nucleare, Sezione di Firenze, Via Sansone 1, 50019 Sesto Fiorentino (FI), Italy}
\affil[24]{ASI -Agenzia Spaziale Italiana, Via del Politecnico snc, 00133 Roma, Italy}
\affil[25]{Istituto Nazionale di Fisica Nucleare, Sezione di Roma "Tor Vergata", Via della Ricerca Scientifica 1, 00133 Roma, Italy}
\affil[26]{Department of Physics and Kavli Institute for Particle Astrophysics and Cosmology, Stanford University, Stanford, California 94305, USA}
\affil[27]{Astronomical Institute of the Czech Academy of Sciences, Bo\v{c}n\'{i}  II 1401/1, 14100 Praha 4, Czech Republic}
\affil[28]{RIKEN Cluster for Pioneering Research, 2-1 Hirosawa, Wako, Saitama 351-0198, Japan}
\affil[29]{California Institute of Technology, Pasadena, CA 91125, USA}
\affil[30]{Yamagata University,1-4-12 Kojirakawa-machi, Yamagata-shi 990-8560, Japan}
\affil[31]{Osaka University, 1-1 Yamadaoka, Suita, Osaka 565-0871, Japan}
\affil[32]{Department of Physics, Faculty of Science and Engineering, Chuo University, 1-13-27 Kasuga, Bunkyo-ku, Tokyo 112-8551, Japan}
\affil[33]{Institute for Astrophysical Research, Boston University, 725 Commonwealth Avenue, Boston, MA 02215, USA}
\affil[34]{Department of Astrophysics, St. Petersburg State University, Universitetsky pr. 28, Petrodvoretz, 198504 St. Petersburg, Russia}
\affil[35]{Physics Department and McDonnell Center for the Space Sciences, Washington University in St. Louis, St. Louis, MO 63130, USA}
\affil[36]{Finnish Centre for Astronomy with ESO, 20014 University of Turku, Finland}
\affil[37]{Universit\'e de Strasbourg, CNRS, Observatoire Astronomique de Strasbourg, UMR 7550, 67000 Strasbourg, France}
\affil[38]{MIT Kavli Institute for Astrophysics and Space Research, Massachusetts Institute of Technology, 77 Massachusetts Avenue, Cambridge, MA 02139, USA}
\affil[39]{Graduate School of Science, Division of Particle and Astrophysical Science, Nagoya University, Furo-cho, Chikusa-ku, Nagoya, Aichi 464-8602, Japan}
\affil[40]{Hiroshima Astrophysical Science Center, Hiroshima University, 1-3-1 Kagamiyama, Higashi-Hiroshima, Hiroshima 739-8526, Japan}
\affil[41]{Department of Physics, The University of Hong Kong, Pokfulam, Hong Kong}
\affil[42]{Center for Astrophysics, Harvard \& Smithsonian, 60 Garden St, Cambridge, MA 02138, USA}
\affil[43]{Max Planck Institute for Astrophysics, Karl-Schwarzschild-Str 1, D-85741 Garching, Germany}
\affil[44]{INAF Osservatorio Astronomico di Brera, Via E. Bianchi 46, 23807 Merate (LC), Italy}
\affil[45]{Dipartimento di Fisica e Astronomia, Universit\`a degli Studi di Padova, Via Marzolo 8, 35131 Padova, Italy}
\affil[46]{Dipartimento di Fisica, Universit\`a degli Studi di Roma "Tor Vergata", Via della Ricerca Scientifica 1, 00133 Roma, Italy}
\affil[47]{Department of Astronomy, University of Maryland, College Park, Maryland 20742, USA}
\affil[48]{Anton Pannekoek Institute for Astronomy \& GRAPPA, University of Amsterdam, Science Park 904, 1098 XH Amsterdam, The Netherlands}
\affil[49]{Guangxi Key Laboratory for Relativistic Astrophysics, School of Physical Science and Technology, Guangxi University, Nanning 530004, China}

\affil[$\ddag$]{Deceased}



\abstract{Using observations of X-ray pulsar Her X-1 by the Imaging X-ray Polarimetry Explorer, we report on a highly significant ($>17\sigma$) detection of the polarization signal from an accreting neutron star. 
The observed degree of the linear polarization of $\sim$10\% is found to be far below theoretical expectations for this object, and stays low throughout the spin cycle of the pulsar.
Both the polarization degree and the angle exhibit variability with pulse phase, which allowed us to measure the pulsar spin position angle 57(2) deg and the magnetic obliquity 12(4) deg, which is an essential step towards detailed modelling of the intrinsic emission of X-ray pulsars.
Combining our results with the optical polarimetric data, we find that the spin axis of the neutron star and the angular momentum of the binary orbit are misaligned by at least $\sim$20 deg, which is a strong argument in support of the models explaining stability of the observed super-orbital variability with the precession of the neutron star.}

\maketitle

\section{Main Text}

X-ray pulsars are strongly magnetized neutron stars powered by accretion from a donor star in binary systems. 
The strong magnetic field funnels the accreting material to the polar caps of the compact object where the energy is released producing the observed pulsed emission as the neutron star rotates.
Her X-1 is the second X-ray pulsar ever discovered \cite{1972ApJ...174L.143T}, one of the few persistent accretion powered pulsars in the sky, and is arguably the most studied object of its type. 
Her X-1/HZ Her is an intermediate mass X-ray binary at a distance of $\sim7$\,kpc \cite{bj21} consisting of a persistently accreting neutron star with the spin period of $\sim1.24$\,s and a B3, $\sim2.2$ solar mass donor star eclipsing the X-ray source every $\sim1.7$\,d as they orbit each other in a nearly circular orbit \cite{1972ApJ...174L.143T,1976ApJ...208..567M}. 
The neutron star has strong magnetic field of $4.5\times10^{12}$\,G, and Her~X-1 is actually the first neutron star where the field was measured directly through the detection of a cyclotron resonance scattering feature in the X-ray spectrum \cite{1978ApJ...219L.105T}. 
Besides the spin and orbital variations, also surprisingly stable $\sim35$\,d super-orbital variability is observed in this system \cite{1973ApJ...184..227G}. Flux variability is thought to be related to obscuration of the compact object by the precessing warped accretion disc at certain precession phases, and is accompanied by regular changes of the pulse profiles. The latter fact motivated a hypothesis that a precession of the accretion disc might be clocked by the neutron star precession via some feedback mechanism \cite{1986ApJ...300L..63T,2009A&A...494.1025S,2013MNRAS.435.1147P}.

The X-ray radiation from Her X-1 was anticipated to be strongly polarized with up to 60--80\% polarization degree (PD) expected in some models \cite{2021MNRAS.501..129C}, so it was chosen as one of the first targets for the Imaging X-ray Polarimetry Explorer (IXPE), a NASA
mission in partnership with the Italian space agency (ASI) equipped with detectors sensitive to linear polarization of the X-rays in the nominal 2--8 keV band.
Here we report on the results of these observations and on the first measurement  of the linear polarization from an accreting neutron star. We also discuss how polarimetry can be used to constrain the basic geometry of the pulsar and test the hypothesis of free precession of the neutron star in this binary system, as well as the challenges it poses for X-ray pulsar emission models.

\begin{figure}[h!]
\includegraphics[width=0.4\textwidth]{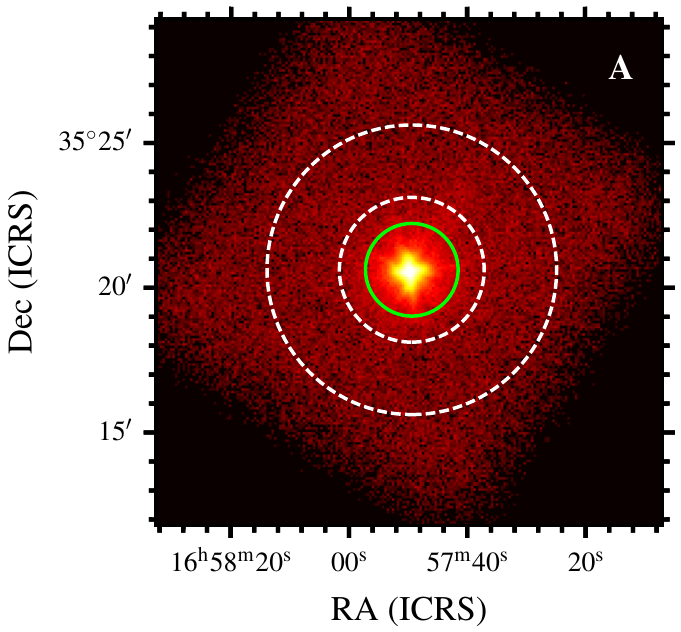}
\includegraphics[width=0.57\textwidth]{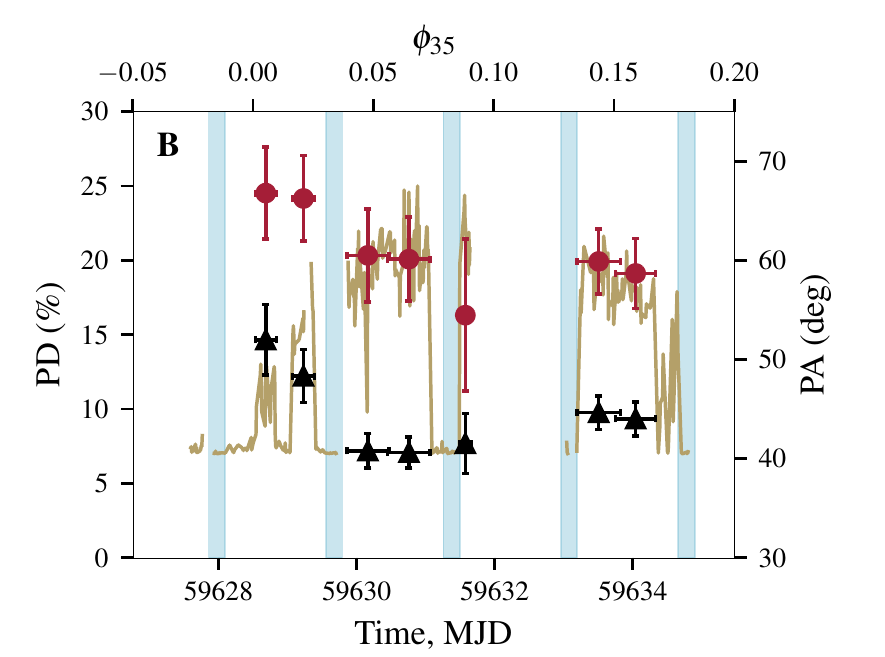}
\caption{\textbf{Overview and evolution of polarization properties of Her~X-1 over the observation.} 
\textbf{(A)} Source (green) and background (white) extraction regions on top of a broadband (2--7\,keV) image of Her~X-1 observed by IXPE (all three detectors combined). 
\textbf{(B)} Evolution of the observed flux from Her X-1  (brown curve), polarization degree (PD, black triangles, left axis) and polarization angle (PA, red circles, right axis) with time (numerical values are listed in Supplementary Table~3). 
The turn-on time MJD~59628.5 is estimated from the IXPE data and the super-orbital period of 34.85\,d is assumed (the corresponding phase is marked at the top axis). The reported values and the uncertainties correspond to the mean values and $1\sigma$ confidence intervals.
The vertical blue stripes show eclipses  by the companion star (eclipses and pre-eclipse dips are excluded from the analysis). 
The error bars correspond to the 68\% confidence level.  
}
\label{fig:timeres}
\end{figure}

The source was observed by IXPE on 2022 February 17--24, at the beginning of the 35~d precession cycle, the so-called ``main-on'' state, as illustrated in Fig.~\ref{fig:timeres}. 
The observation started while the pulsar was still obscured by the outer edge of the warped and tilted accretion disc \cite{2003MNRAS.342..446L,2006AstL...32..804K} and continued throughout the first part of the main-on state where the neutron star emerges from behind the accretion disc and becomes visible directly \cite{2021Univ....7..160L}. 
IXPE had, therefore, a direct and clear view of the neutron star through most of the observation except for brief periods when the pulsar was eclipsed by the donor star, and the so-called ``pre-eclipse'' dips, associated with obscuration by the outer disc regions disturbed by the interaction with the accretion stream from the donor star \cite{2012MNRAS.425....8I} or by the gas stream itself \cite{1999A&A...348..917S}. 
The data taken during the eclipses of the pulsar and during periods of strong absorption were excluded from the analysis.
This resulted in a total effective exposure time of $\sim150$\,ks suitable for polarimetric and spectro-polarimetric analysis based on the formalism outlined by \cite{2015APh....64...40K} and \cite{2017ApJ...838...72S} and standard for all IXPE observations up to now, which is described in detail in Methods.

\subsection*{Results}

\begin{figure}[h!]
\centering\includegraphics[width=0.8\textwidth]{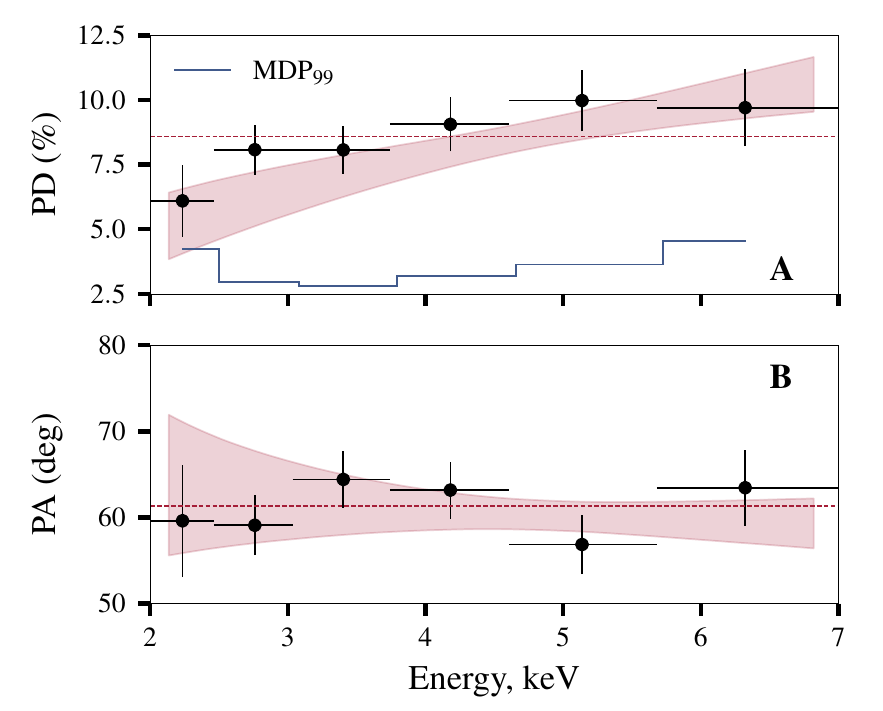}
\caption{\textbf{Energy dependence of the polarization in Her~X-1.} 
\textbf{(A)} Pulse-phase averaged PD and \textbf{(B)} PA as a function of photon energy estimated using the formalism of ref. \cite{2015APh....64...40K} are shown with the black circles. 
The y-axis error bars correspond to $1\sigma$, while the x-axis error bars reflect the width of the energy bins used for binned analysis. 
The blue line shows the estimated minimal detectable polarization at the 99\% confidence level for each bin. 
The shaded regions corresponds to $1\sigma$ confidence interval for spectro-polarimetric analysis with the \texttt{polpow} model. The dashed horizontal lines indicate average values of the PD and PA over the full energy band.
}
\label{fig:energy_resolved}
\end{figure}

We started the analysis by looking at the phase-averaged polarization of the emission from Her~X-1, using all photons collected throughout the observation in the broad 2--7\,keV energy band, ignoring the 7--8\,keV band due a higher background and remaining calibration uncertainties.
We detect a highly significant and well constrained polarization signal, with a polarization degree (PD) of $8.6\pm0.5\%$ and polarization angle (PA, measured from north to east) of $62\degr\pm2\degr$ (all uncertainties are quoted at $1\sigma$ confidence level unless stated otherwise). 
The measured PD is significantly lower than the predicted 60--80\% for the source \cite{2021MNRAS.501..129C}, which opens the way for new theoretical investigations as we discuss below. We emphasise that the unexpectedly low polarization is clearly intrinsic to the radiation emerging from the pulsar, and cannot be explained with the signal being depolarized on its way from the pulsar to the observer, e.g., by scattering in the accretion flow or accretion disc atmosphere. 
Indeed, as already mentioned, the source is expected to be observed directly throughout most of the observation, and moreover, the PD appears to be minimal at the peak of the main-on where the flux is maximal and thus the amount of scattering material minimal as illustrated in Fig.~\ref{fig:timeres}.
As the next step, we investigated the dependence of the polarization properties on photon energy. 
We find that both the PD and PA appear to be energy independent (see Fig.~\ref{fig:energy_resolved} and Extended Data Figure~1), with only an indication at $\sim2\sigma$ confidence level (see Methods) for the PD increasing towards higher energies. 
We continue, therefore, discussing only the energy-averaged polarization properties within the relatively narrow energy band covered by IXPE.

\begin{figure}[h!]
\centering \includegraphics[width=1\textwidth]{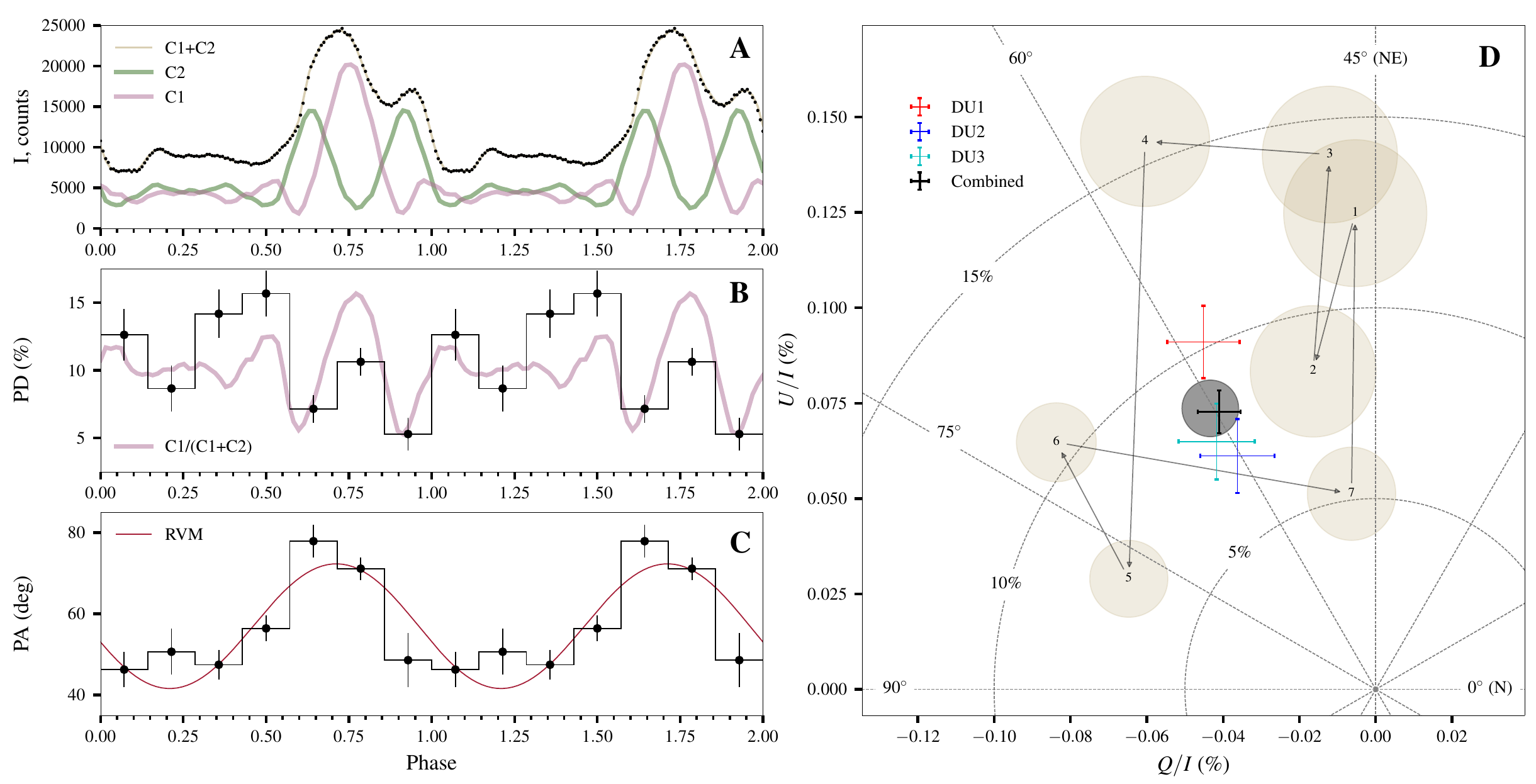}
\caption{\textbf{Pulse-phase dependence of the polarization properties in Her~X-1.}  
\textbf{(A)} Observed pulse profile in the 2--7\,keV energy range (counts per 1/128 phase interval) and its decomposition into single-pole pulse profiles labelled as C1 and C2 \cite{2000ApJ...529..968B}. 
\textbf{(B)} PD and \textbf{(C)} PA estimated from the spectro-polarimetric fit are shown as a function of pulse phase with black circles. 
The violet line in panel (B) shows relative contributions of the main pole (C1 component, which dominates the main peak) to the total flux, and the red line in panel (C) shows the best-fitting approximation for the PA with the rotating vector model. 
\textbf{(D)} Normalized Stokes parameters $Q/I$ and $U/I$ are shown for each phase bin with brown ellipses representing 1$\sigma$ confidence regions for Stokes parameters (numbers indicate bin number in panels A and B from left to right). 
The black circle shows the Stokes parameters for the pulse-phase averaged analysis based on the spectro-polarimetric fit.
The results for the unbinned analysis \cite{2015APh....64...40K}  for individual detector units and combining the three detectors are shown with coloured error bars. 
In all panels points correspond to the mean values and the error bars correspond to $1\sigma$ confidence level.}
\label{fig:phaseres}
\end{figure}

Pulsar geometry can only be constrained through analysis of the pulse-phase dependence of the polarization properties and we do indeed observe strong and highly significant variations of the polarization properties with the spin phase, as illustrated in Fig.~\ref{fig:phaseres}. We note that the PD remains well below expectations for all pulse phases, never exceeding $\sim15$\%, i.e. not dramatically higher than the phase-averaged value.  The phase dependence of the observed PD is rather complex whereas PA shows simpler, roughly sinusoidal dependence. 
The observed spin-phase dependence of the PA can be interpreted within the basic assumptions of X-ray pulsar modelling. 
In fact, photons coming  from different parts of the emission region are expected to substantially align with the magnetic field as they propagate in the highly magnetized plasma surrounding the X-ray pulsar. 
Vacuum birefringence causes the polarized radiation in the magnetosphere to propagate in the normal, ordinary (O) and extraordinary (X), modes which represent oscillations of the electric field parallel and perpendicular to the plane formed by the local magnetic field and the photon momentum \cite{1978SvAL....4..117G,1979JETP...49..741P}, and propagation in the normal modes continues within the so-called polarization limiting radius \cite{heylsh00}. 
This radius is estimated to be about thirty stellar radii for typical X-ray pulsars \cite{2018Galax...6...76H}, and at such distances, the field is expected to be dominated by the dipole component. 
The polarization measured at the telescope is expected, therefore, to be either parallel or perpendicular to the instantaneous projection of the  magnetic dipole axis of the star onto the plane of the sky. In this scenario the variation of the PA with phase is a purely geometrical effect and therefore it is not related at all to changes of the PD or flux.

\begin{figure}[h!] 
\centering \includegraphics[width=0.9\textwidth]{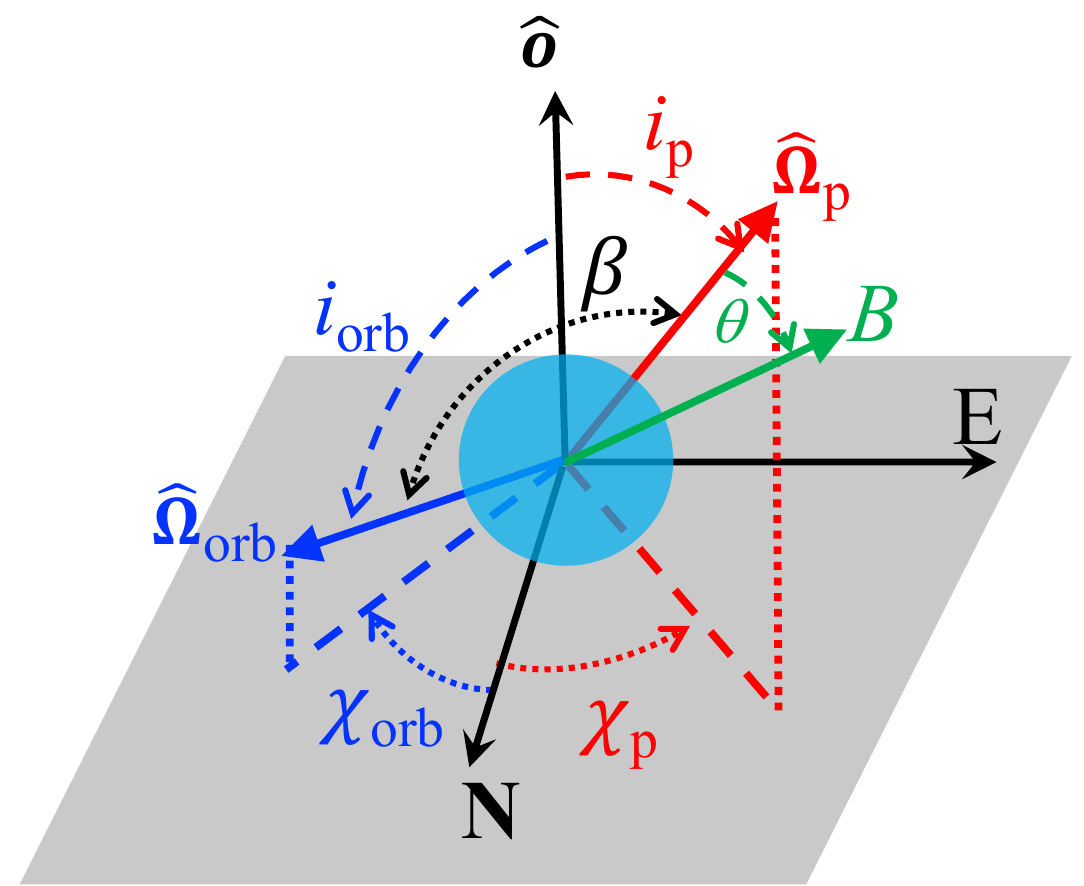}
\caption{\textbf{Geometry of the system from the observer's perspective.} 
The grey plane is the plane of the sky, labelled with north and east axes, perpendicular to the line of sight towards the observer $\unit{o}$. 
The angles between the line of sight and the vectors of the pulsar spin $\unitvect{\Omega}_{\rm p}$ and the orbital angular momentum $\unitvect{\Omega}_{\rm orb}$ are the inclinations  $i_{\rm orb}$  and $i_{\rm p}$. 
The corresponding position angles $\chi_{\rm p}$ and $\chi_{\rm orb}$ are the azimuthal angles of the spin vectors projected onto the sky, measured from north to east. 
The misalignment angle $\beta$ is defined as the angle between $\unitvect{\Omega}_{\rm p}$ and $\unitvect{\Omega}_{\rm orb}$. 
The magnetic obliquity $\theta$ is the angle between magnetic dipole and the rotational axis. 
}
\label{fig:geometry}
\end{figure}

Based on these considerations, we can constrain the pulsar geometry by modelling the pulse-phase dependence of the PA with the rotating vector model (RVM) \cite{RC69}.
If the position angle (measured from north to east) of the pulsar angular momentum is $\chi_{\rm p}$ and the pulsar radiation is dominated by the ordinary O-mode, then the variations of the X-ray PA with the pulsar phase $\phi$ can be described by the expression  \cite{Poutanen20RVM}   
\begin{equation} \label{eq:pa_rvm}
\tan (\mbox{PA}-\chi_{\rm p})= \frac{-\sin \theta\ \sin (\phi-\phi_0)}
{\sin i_{\rm p}\ \cos \theta  - \cos i_{\rm p}\ \sin \theta\  \cos (\phi-\phi_0) } ,
\end{equation} 
where $i_{\rm p}$ is the inclination of the neutron star angular momentum to the line of sight (defined in the interval $[0\degr,180\degr]$), $\theta$ is the inclination of the magnetic dipole to the spin axis (i.e. the magnetic obliquity), 
$\phi_0$ is the phase of the light curve when the spot is closest to the observer (see Fig.~\ref{fig:geometry} for geometry). 

\begin{figure}[h!]
\centering \includegraphics[width=1.0\textwidth]{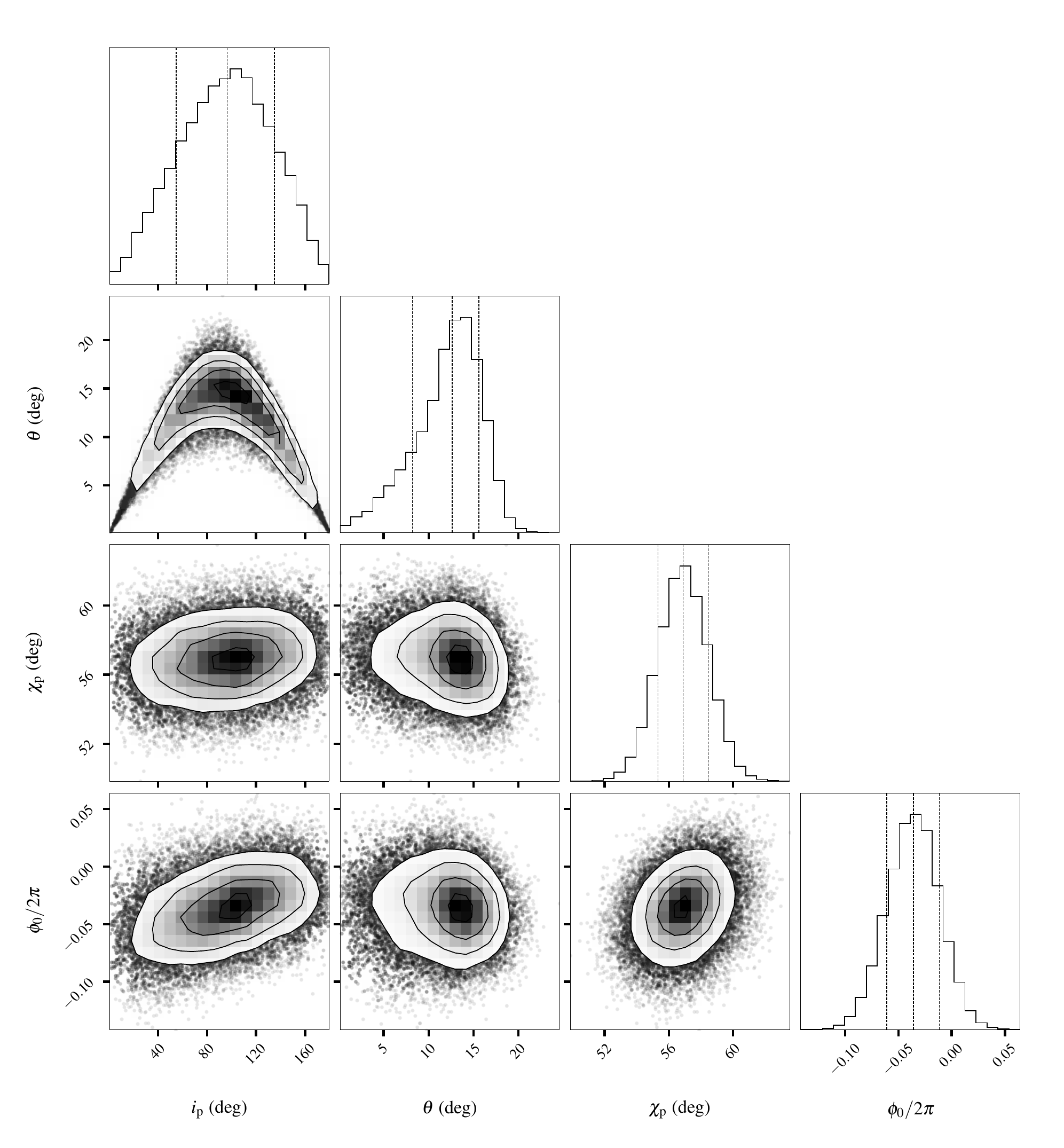}
\caption{\textbf{Posterior distribution corner plot for the RVM fit of the PA phase dependence}. 
The contours correspond to two-dimensional 1, 2, and 3$\sigma$ levels. 
\label{fig:corner_rvm}} 
\end{figure}

Using Bayesian inference code BXA \cite{Buchner14}, we fit the PA data from Fig.~\ref{fig:phaseres} using RVM with four free parameters ($\chi_{\rm p}$, $\theta$, $i_{\rm p}$, and $\phi_0$).
We assume flat priors for all parameters: $\chi_{\rm p}\in [-90\degr,90\degr]$, $\theta \in [0\degr,90\degr]$, $i_{\rm p}\in[0\degr,180\degr]$, and $\phi_0/(2\pi)\in[-0.5,0.5]$. 
The resulting posterior distributions are shown in Fig.~\ref{fig:corner_rvm}. 
The magnetic obliquity and the pulsar position angle are both well constrained $\theta=12\fdg1\pm 3\fdg7$ and $\chi_{\rm p}=\chi_{\rm p,*}=56\fdg9\pm 1\fdg6$. 
Because only the orientation of the polarization plane can be measured, the polarimetric data cannot distinguish between oppositely directed pulsar spins. Therefore, there exists another solution with the oppositely directed spin: $\chi_{\rm p}=\chi_{\rm p,*}\pm180\degr$. 
If radiation escaping from the pulsar is polarized perpendicular to the magnetic field direction (i.e. in the X-mode), then the position angle of the pulsar spin can have two possible values: $\chi_{\rm p}=\chi_{\rm p,*}\pm 90\degr=146\fdg9\pm 1\fdg6$ or $-33\fdg1\pm 1\fdg6$, respectively. 
Other angles (in particular, $\theta$) are not affected by the spin direction or  uncertainty in intrinsic polarization of radiation escaping from the surface.  
We emphasise that the value for $\theta$ is in excellent agreement with the indirect estimates obtained from the modelling of the observed pulse profile shape \cite{2000ApJ...529..968B}.
This both lends support to our assumption that the PA at least approximately follows the RVM model and lends some credibility to the aforementioned modelling of the pulse profile shapes. 
It is important to emphasise that all previous estimates of the magnetic co-latitude were based on indirect arguments whereas our measurement is direct, and the position angle of the pulsar's rotation axis on the sky is measured for the first time.

On the other hand, the X-ray polarimetry alone does not allow us to obtain meaningful constraints on the pulsar inclination although our measurement is still fully consistent with the independent estimates of the binary orbit inclination \cite{Drissen86}. Indeed, pulsar inclination has a rather large uncertainty, $i_{\rm p}=95\degr\pm37\degr$, with the posterior probability distribution extending from 0\degr\ all the way up to 180\degr,  that can be approximated by the function
\begin{equation}\label{eq:distr_inclp}
\frac{dp}{di_{\rm p}} \propto 
\left\{ \begin{array}{ll}
\displaystyle   \sin^{1.5}(90\degr\ i_{\rm p}/i_{\rm peak}), 
& i_{\rm p} \leq i_{\rm peak} \\
\displaystyle  \sin^{1.4}[90\degr\ (2i_{\rm peak}-i_{\rm p}-180\degr)/(i_{\rm peak}-180\degr)],
& i_{\rm p} > i_{\rm peak},
\end{array} \right.
\end{equation} 
where $i_{\rm peak}=97\degr$ is the angle where the distribution peaks. This value is indeed consistent with literature estimates for the orbital inclination of $i_{\rm orb}\sim80\degr-90\degr$ \cite{1997MNRAS.288...43R,leahy14}.

Considering that free precession of the neutron star has been previously suggested to explain stability of the 35\,d precession cycle \cite{1986ApJ...300L..63T,2009A&A...494.1025S,2013MNRAS.435.1147P,2022MNRAS.513.3359K}, it is, however, still interesting to test whether spin axis of the pulsar and orbital angular momentum are aligned. This can be done despite the fact that inclination of the pulsar with respect to the line of sight is poorly constrained by X-ray polarimetry alone if orientation of the orbital plane on the sky is known. Such constraints can be obtained from the optical polarimetric observations of Her~X-1 over its orbital period \cite{1991A&A...244L..41E} and assuming that optical polarization results from scattering by an optically thin material corotating with the system as seen by {eRosita} \cite{2021A&A...648A..39S}.  
To do that we started by fitting the phase curves of the normalized Stokes parameters digitalized from Fig.~1 in \cite{1991A&A...244L..41E} with the Fourier series 
\begin{equation}\label{eq:Fourier}
\begin{array}{ll}
q  & = q_0 + q_1\cos{\varphi} + q_2\sin{\varphi} + q_3\cos{2\varphi} + q_4\sin{2\varphi}, \\
u  & = u_0 + u_1\cos{\varphi} + u_2\sin{\varphi} + u_3\cos{2\varphi} + u_4\sin{2\varphi} ,
\end{array}
\end{equation}
where $\varphi$ is the orbital phase. 
If the polarization is produced by Thomson scattering in an optically thin medium co-rotating with the system, the orbital orientation can be obtained from the Fourier coefficients \cite{Brown1978}.  
The best-fitting Fourier coefficients and their errors obtained by us are given in Supplementary Table~4 and are close to those reported in \cite{1991A&A...244L..41E}.
These coefficients can be used to derive the inclination $i_{\rm orb}$ of the binary orbit and the position angle $\chi_{\rm orb}$ of the projection of the orbital axis \cite{Drissen86,Kravtsov20}:
\begin{equation}\label{eq:Drissen_i}
\left(\frac{1 - \cos{i_{\rm orb}}}{1 + \cos{i_{\rm orb}}}\right)^4 = \frac{(u_3 + q_4)^2 + (u_4 - q_3)^2}{(u_4 + q_3)^2 + (u_3 - q_4)^2} , 
\end{equation}
\begin{equation}\label{eq:Drissen_omega}
\tan{(2\chi_{\rm orb})} = \frac{A + B}{C + D},\\
\end{equation}
where
\begin{equation}
\begin{array}{ll}
A &= \displaystyle\frac{u_4 - q_3}{(1 -\cos{i_{\rm orb}})^2},\quad B = \frac{u_4 + q_3}{(1 +\cos{i_{\rm orb}})^2} ,\\
C &= \displaystyle\frac{q_4 - u_3}{(1 +\cos{i_{\rm orb}})^2},\quad D = \frac{u_3 + q_4}{(1 -\cos{i_{\rm orb}})^2}. 
\end{array}
\end{equation}
These formulae give us $i_{\rm orb} =100\fdg4\pm 4\fdg9$ and $\chi_{\rm orb}=\chi_{\rm orb,*}=28\fdg9\pm5\fdg9$ (or $\chi_{\rm orb}=\chi_{\rm orb,*}-180\degr= -151\fdg1\pm5\fdg9$ which is equally acceptable since only the orientation of the polarization plane can be measured). The final values for all geometrical parameters of the system including constraints from X-ray and optical polarimetry are summarised in Table~\ref{tab:geometry}.
The obtained orbital inclination is larger than 90\degr, which might appear to be at odds with independent inclination estimates quoted above. 
We note, however, that these estimates are based on modelling of the donor star radius from optical spectroscopy and X-ray eclipses, and cannot distinguish between clockwise and counter-clockwise rotation (i.e. between inclinations  $i_{\rm orb}<90\degr$ and $180\degr-i_{\rm orb}$). 
In particular, the estimates listed in Table~8 of ref. \cite{leahy14} seem to favour inclinations in the range $i_{\rm orb}\sim80\degr-83\degr$ or $180\degr-i_{\rm orb}\sim97\degr-100\degr$ 
for the distance range of 6.5--7.5\,kpc, estimated from Gaia EDR3 data \cite{bj21}.
This implies that our estimate is fully consistent with the literature values, and that the binary is rotating clockwise on the sky. We emphasise that this is a new result which can only be obtained from polarimetry, in this case, in the optical band.

Using constraints on the 3D orientation of the pulsar and the orbit, we now can obtain the  misalignment angle $\beta$ between the pulsar and the orbital angular momenta:
\begin{equation} 
    \cos\beta = \cos i_{\rm p} \cos i_{\rm orb} + \sin i_{\rm p} \sin i_{\rm orb} \cos \Delta,
\end{equation} 
where $\Delta=\chi_{\rm p}-\chi_{\rm orb}$ is the difference between the position angles of the pulsar spin vector and the orbital angular momentum  (the geometry is illustrated in Fig.~\ref{fig:geometry}). 
The parameters we use are given in Table~\ref{tab:geometry}. 
Assuming normal distributions for $\chi_{\rm p}$ and $\chi_{\rm orb}$ with the corresponding $1\sigma$ errors obtained above, a normal distribution for $i_{\rm orb}$ from the optical polarimetry, and the posterior distribution for $i_{\rm p}$ given by Equation~(\ref{eq:distr_inclp}), we make 
Monte-Carlo simulations to obtain a probability distribution for $\beta$, which is shown in Extended Data Figure 2 and listed in Supplementary Table~5. 
For radiation in the O-mode (when $\chi_{\rm p}=\chi_{\rm p,*}=56\fdg9\pm 1\fdg6$ and taking $\chi_{\rm orb}=28\fdg9\pm5\fdg9$),  we get the smallest misalignment $\beta$ with the distribution peaking at $\sim$30\degr\ and the lower limit being $\sim$20\degr\ at the 90\% confidence level (see Extended Data Figure 2A). 
If $\chi_{\rm p}=\chi_{\rm p,*}
\pm180\degr$ (or $\chi_{\rm orb}=\chi_{\rm orb,*}\pm180\degr$), the misalignment is much larger, with $\beta$ peaking at $145\degr$ (Extended Data Figure 2B). 
For the X-mode polarization, $\chi_{\rm p}=\chi_{\rm p,*}\pm 90\degr$, $\beta$ peaks at $\sim$115\degr\ or $\sim$65\degr\ (Extended Data Figure 2C,D). These results are practically unaffected by the exact form of the distribution of $i_{\rm p}$ and imply that the spin axis of the neutron star during the observation is inclined with respect to the orbital spin by at least 20\degr\ and possibly by as much as $\sim$160\degr\  (see Extended Data Figure 2). 
We note that low angular momentum of the neutron star implies that accretion torques are expected to align its spin with the orbital angular momentum on a relatively short timescale \cite{1991ApJ...378..696P,Biryukov21}, so naively one could expect spin of the pulsar and orbital angular momentum to be aligned. This is, however, apparently not the case.

\subsection*{Discussion}
Meaningful interpretation of the observed variation of the PD with pulse phase is only if the spectra, the pulse profiles and, now, the observed polarization properties of X-ray pulsars are consistently explained. The observed low degree of polarization in Her~X-1 came as a surprise and is inconsistent with predictions, and therefore, can not be interpreted in framework of existing models. One could think, however, of several potential scenarios explaining observed low PD. For instance, radiative transfer in the magnetized plasma within the emission region with specific temperature structure of the neutron star atmosphere can be responsible for observed low PD (see Methods). 
Propagation of the initially polarized X-rays through the magnetosphere can also result in depolarization due to QED effects \cite{2021MNRAS.501..109C}. 
In either scenario, averaging over wider pulse phase intervals or over energy can be expected to reduce the observed PD. Finally, we likely observe emission from both poles of the neutron star combined at least at some pulse  phases \cite{2000ApJ...529..968B}. Each of the poles could have different polarization properties since both are observed from different angles at a given pulse phase, and, therefore, mixing the two can reduce the observed PD  (Supplementary Figure 2C). 
Indeed, modelling of the evolution of the complex observed pulse profile shape over the 35-d cycle \cite{2013MNRAS.435.1147P} suggests multiple emission regions likely related to non-dipolar structure of magnetic field close to the neutron star surface \cite{2013MNRAS.435.1147P,2022MNRAS.515..571M}.
We note that there is indeed an apparent connection between the observed variations and estimated relative contribution of the pole dominating the main peak of the pulse \cite{2000ApJ...529..968B}, as illustrated in Fig.~\ref{fig:phaseres}B. 
This might suggest that mixing of the emission from different poles might be at least partly responsible for the observed low PD, and it also suggests that the decomposition of the observed pulse profile to single pole components obtained by \cite{2000ApJ...529..968B} is probably not far from reality. 
The PD remains, however, low even during the peak where emission is dominated by a single pole. The contribution of the two poles is thus not the only reason for the observed low PD, and probably a combination of several mechanisms is at work. In general, it is clear that a full interpretation of the observed polarization properties of Her~X-1 (and other X-ray pulsars) and a full assessment on the scenarios outlined above, requires a deeper understanding of the accretion physics and the emission mechanisms in these objects.
This includes the pulse shape, the broad-band energy spectrum and its variations with spin and precession phase, the periodic and secular variations in its cyclotron absorption feature and, of course, polarization properties. 
Up to now there is no theoretical model explaining all these observables, and particularly polarization. The observed low PD, therefore, already puts strong constraints on the possible emission mechanisms at play in accretion-powered pulsars, and constitutes a valuable input for theoretical modelling of emission from accreting magnetized neutron stars. 

The polarimetric observations reported in this work also provide previously unavailable information on the geometry of the source, in particular, basic information on orientation of the pulsar geometry including magnetic co-latitude and orientation with respect to observer and to the orbit of the binary system. 
In particular, we find evidence of a misalignment between the spin axis of the pulsar and the orbital angular momentum. The reason for the observed misalignment is unclear, but it could be associated, for instance, with extra torques imposed on the neutron star by the warped accretion disc or free precession of the neutron star \cite{2013MNRAS.435.1147P}.  
In particular in the latter case, the interaction of the inner disc regions with magnetosphere of a precessing neutron star can greatly diminish or stop altogether secular spin-orbital alignment \cite{2013MNRAS.435.1147P}.
We note that expected alignment was one of the key arguments \cite{1991ApJ...378..696P} against free precession model, and IXPE results invalidate it. It is clear that for a precessing neutron star one can anticipate evolution of the magnetic obliquity $\theta$ with the phase of the 35-d cycle  \cite{1986ApJ...300L..63T} resulting in changing of the amplitude of the PA variations with the spin phase. 
Current observations only cover a small fraction of the 35-d cycle and do not allow to test this hypothesis. 
Deeper observations covering a larger fraction of the cycle would be, however, required to characterise this variability quantitatively and unambiguously prove the hypothesis of the neutron star precession in this system. Furthermore, new high-precision optical polarimetric observations covering different phases of the super-orbital cycle would be useful to confirm the orbital orientation. 
Nevertheless, the obtained constraints on misalignment of the pulsar spin with the orbital angular momentum already represent a strong argument supporting the hypothesis of neutron star precession in the system.
This information can only be obtained by means of polarimetric observations now accessible also in the X-ray band. Our results illustrate the power of X-ray polarimetry for studies of accreting neutron stars, and open a new perspective on these long-known, yet still mysterious objects.

\section{Methods}
\label{sec:methods}
\subsection*{Analysis of IXPE data}
IXPE includes three co-aligned X-ray telescopes, each comprising an X-ray mirror assembly (NASA-furnished), and linear polarization-sensitive pixelated Gas Pixel Detectors (GPDs, ASI-furnished) to provide imaging polarimetry over a nominal 2--8~keV band. A complete description of the hardware and its performance is given in \cite{Weisskopf2022, Soffitta2021, Baldini2021}.  
The GPDs are, in essence, pixelated proportional counters, which allow to recover the direction for each primary photo-electron ejected upon the interaction of an incident photon with the detector medium. 
This direction and the track length carry information about the direction of electromagnetic field oscillation associated with each individual photon, and thus could be used to recover polarization properties (i.e. the Stokes parameters) for an astrophysical source through analysis of the distribution of track directions for all photons from the source. 
The amplitude of variation of the track angles for a 100\% polarized source is described by the energy-dependent modulation factor.
The values and the energy dependence of the modulation factor were calibrated both on ground and continuously monitored in space, and they are taken into account when modelling the polarization as described below. 

IXPE data telemetered to the ground stations in Malindi (primary) and in Singapore (secondary) are transmitted to the Mission Operations Center (MOC) at the Laboratory for Atmospheric and Space Physics (University of Colorado) and then to the Science Operations Center (SOC) at the NASA Marshall Space Flight Center. 
Using the software developed jointly by the NASA and the ASI, the SOC processes science and relevant engineering and ancillary data, to produce the data products that are archived at the High-Energy Astrophysics Science Archive Research Center (HEASARC) at the NASA Goddard Space Flight Center, for use by the international astrophysics community. 
IXPE data are distributed in a lower level format (L1), where relevant information about event tracks are reported, and also in a higher level format (L2), where several corrections have been applied and only the main properties of the reconstructed events are reported.
In particular, in the L2 the photon energy is obtained after corrections for temperature and gain effects. Further corrections for the gain effects are applied using the data from the on-board calibration sources acquired during the observation. 
The imaging information in L2 is obtained from the L1 after correcting for dithering of the spacecraft pointing and orbital thermally-induced motion of the boom that separates the optics from the detectors. The L2 data were then screened and processed using the current version of the \textsc{HEASOFT} software and calibration files.

The data reduction consists of the following main steps. 
The track images are first processed to separate the signal from electronic noise and then a custom algorithm is applied to derive the characteristics of the event, that are, the direction of the photoelectron emission, the energy, the arrival time and the direction of the incoming photon. 
The subsequent steps are to calibrate both the energy and the response to polarization, and to filter bad events and time intervals in which the source was occulted by the Earth or there were pointing inaccuracies, etc. 
 
After initial processing, various selection criteria may be imposed for detected photons. 
Those can include the energy (to study energy dependence of the polarization properties), the arrival time, the pulse or the orbital phase, or position on the detector (to study spatial dependence of the polarization properties in extended sources or to discriminate between source and background photons for point sources).  
On the selected event ensemble, the last step is to normalize the measured response to polarization by the modulation factor. 

Analysis of polarization is carried out with two different approaches. 
The first one, based on the unbinned formalism presented in \cite{2015APh....68...45K}, is implemented in the IXPE collaboration software suite \textsc{ixpebossim} \cite{Baldini2022}. 
The other method relies on the procedure presented in \cite{2017ApJ...838...72S}, and it is based on the generation of the Stokes spectra, which are then fitted with standard spectral-fitting software, like \textsc{xspec} \cite{Arnaud1996}. 
The proper instrument response functions are provided by the IXPE Team as a part of the IXPE calibration database released on 2022 March 14 and available at HEASARC archive (\url{https://heasarc.gsfc.nasa.gov}). 
All values reported below are based on the spectro-polarimetric fits of the Stokes spectra unless stated otherwise. 
The uncertainties are estimated using a Markov chain Monte Carlo (\texttt{mcmc}) method for respective parameter from spectro-polarimetric fits.

\subsubsection*{Pulse-phase averaged analysis}

As a first step, we investigated the time-averaged polarization from the pulsar.
The Stokes parameters are obtained from the L1 data using the unbinned approach of \cite{2015APh....68...45K} and the spurious modulation is removed following the approach of \cite{Rankin_2022}. 
The Stokes parameters in the L2 data are distributed with weights obtained following the procedure from \cite{Di_Marco_2022}, which can be used to perform a weighted analysis improving the sensitivity for faint sources. 
Considering the low background level and the high number of source counts in the case of Her~X-1, we do not use the weighted approach for the final results reported.
We performed, however, both weighted and unweighted analyses and found compatible results. 

The source and background photons were extracted from a circular (radius of 1\farcmin 6) and annular (with inner and outer radii of 2\farcmin5 and 5\arcmin, respectively) regions centred at the source. 
The extraction radii were chosen to select the source with a proper margin; the background was later removed by subtracting its Stokes parameters, re-scaled for the appropriate extraction area, from those of the source.
The average values of the Stokes parameters, and corresponding polarization degree (PD) and polarization angle (PA), were then estimated in a single 2--7\,keV energy band and in four sub-bins covering the same energy range. 
Note that we conservatively ignored energies in 7--8\,keV energy range to avoid potential systematic effects associated with the remaining energy scale uncertainties (which can be expected to have largest effect around the energies where effective area drops abruptly, i.e. around 8\,keV) and uncertainties in the alignment of the optical axis at this stage of the mission, which affects vignetting correction (which is again strongest at highest energies). 
We emphasise, however, that these effects mostly affect spectral analysis (i.e. the best-fitting parameters of the spectral model), and the polarimetric results are not affected. 

In addition to the binned analysis, we have also conducted spectro-polarimetric modelling of the same data-set. 
In particular, the Stokes spectra were extracted for each detector unit and modelled simultaneously using absorbed \textsc{nthcomp} model \cite{1999MNRAS.309..561Z} for intensity spectra in combination with either \textsc{polconst} or \textsc{polpow} polarimetric models. 
The \textsc{nthcomp} model describes a Comptonized spectrum from seed photons of a characteristic temperature $T_{\rm bb, {comp}}$ (defining the low energy rollover) by electrons with temperature $T_{\rm e, {comp}}$ (defining the high energy rollover). 
Instead of the Thomson optical depth this model is parametrized by the power-law index $\Gamma_{\rm comp}$, because the Comptonization spectrum for non-relativistic electron temperatures is well described by a power law  between the photon seed energies and the cutoff energy related to the electron temperature. 
This model is often used to describe the spectra of X-ray pulsars. 
The model normalization at 1\,keV, $A_{\rm comp}$, and cross-normalization constants defining relative normalization of {IXPE} detector units two and three relative to the first one, $C_{\rm DU2}$ and $C_{\rm DU3}$, were also considered as free parameters.

We emphasise that \textsc{nthcomp} is a purely phenomenological model and physical interpretation of the best-fitting values is not trivial as the model is actually not designed to describe the spectra of X-ray pulsars.
The spectrum of Her~X-1 is known to be more complex than given by this model (e.g. there is a blackbody-like component with $kT\sim0.1-0.3$\,keV and a cyclotron absorption line), but within the {IXPE} band the spectrum is well described by this simplified model. 
In fact, the phase-averaged spectrum can even be approximated with a single power law, but this does not apply to all phase bins, hence our choice of the next simplest model. 
We verified, however, that the choice of the intensity continuum model does not significantly affect any of the polarimetric measurements (as is also justified by the agreement between the binned analysis and the spectro-polarimetric analysis results). 

It is worth noting that at the time of the Her~X-1 observation, the IXPE telescope axes were slightly offset with respect to the pointing direction, and that there were uncertainties in modelling of the boom motion during the observation. 
This caused an additional vignetting with an impact on the effective area calibration and then on the spectral analysis. 
However, this has no impact on the measured dependence of the polarization on energy because the polarization is estimated after normalization of the Stokes parameters \textit{U} and \textit{Q} to the source flux, which cancels out the systematics related to the effective area. 
This is also confirmed by the analysis presented in Fig.~\ref{fig:timeres}, Extended Data Figure 1 and Supplementary Table~1.
We emphasise a good agreement between the individual detectors and the two independent modelling approaches.

The polarization properties appear to be only weakly dependent on energy, although there is an indication of increase of the PD with energy. 
Indeed, although there appears to be a systematic increase of the PD towards higher energies, and the value of Pearson correlation coefficient between PD and energy of $\sim0.86$ suggests moderate degree of correlation, the values in individual bins, except the first one, are consistent with the average value, as illustrated in Fig.~\ref{fig:energy_resolved}. 
An alternative approach to assess the significance for such energy dependence is to compare the results of the spectro-polarimetric fits for models when polarization is assumed constant to those where it is energy-dependent, which are summarised in Supplementary Table~1.
Indeed, the model where constant polarization is assumed yields slightly worse fit statistics, but a lower Bayesian information criterion (BIC) score \cite{bic_ref}, which makes it statistically preferred. 
Similar conclusion can be drawn based on the estimated significance of the deviation of the power-law index, characterising the PD dependence on energy PD$(E)\propto E^{-\Gamma_{\rm PD}}$, from zero, which is estimated at $\Gamma_{\rm PD}=-0.46\pm0.20$. 
It deviates from zero at the confidence level of only $\sim98$\%, i.e. at $\sim2\sigma$. 
The power-law index characterizing the dependence of the PA is estimated as $\Gamma_{\rm PA}=0.04\pm0.10$, which is consistent with zero. 
We conclude, therefore, that there is no strong dependence of the polarization properties on energy, although there is an indication that the PD might actually increase with energy.

\subsubsection*{Pulse-phase and time-resolved analysis}

In order to investigate the polarization properties as a function of the spin phase, we obtained a timing solution for the pulsar.
As a first step, the arrival times of all events were corrected to the Solar system barycentre reference frame using the \textit{barycorr} task, and then were corrected for effects of motion within binary system using ephemerides by \cite{2009A&A...500..883S}. 
After that, a Lomb-Scargle \cite{Lomb76,Scargle82} periodogram was constructed to estimate the approximate value of the spin period and to obtain a template pulse profile which was used to estimate the residual phase delays and the pulse arrival times for observation segments by cross-correlation with the template (we considered continuous segments separated by at least 1\,ks gaps as independent). 
The obtained pulse arrival times $t_n$ were then used to obtain the final estimate of the spin period $p_{\rm spin}=1.2377093(2)$\,s using the phase connection technique. 
In particular, we found that the observed arrival times were fully consistent with a constant period, i.e. $t_n=t_0+n\times p_{\rm spin}$ as illustrated in Supplementary Figure 1. 
It is important to emphasise that no appreciable evolution of the pulse profile shape occurs during the observation as illustrated in Supplementary Figure 1, and expected on the basis of previous observations of the source at a similar phase of the precession cycle \cite{2001ESASP.459..309K}. 
This allows us to use all the available data and achieve a sufficient sensitivity also in the individual phase bins. The observed pulsed fraction in the 2--7\,keV band, defined through the maximum and minimum fluxes as $f=(F_{\max}-F_{\min})/(F_{\max}+F_{\min})$, is $\sim55$\%.

Based on the available counting statistics and known instrument sensitivity, seven phase bins were then defined as shown in Fig.~\ref{fig:phaseres}. 
The Stokes spectra ($I/Q/U$), and binned polarization cubes, were then extracted individually for each of the phase bins using \textsc{ixpeobssim} package \cite{Baldini2022}. 
The background was assumed to be constant for all bins (which is justified since minor variations of the background rate during the observations are averaged out when folded with the spin period of the source). 
We used, therefore, Stokes spectra extracted for the entire observation as a background estimate in the phase-resolved analysis (after accounting for difference in the exposure).
The extracted spectra were then modelled with the same model as the pulse-phase averaged spectra to derive the PD and PA using the \texttt{polconst} model. 
The final values and uncertainties were estimated based on \textit{mcmc} chains produced using the \texttt{chain} command in \textsc{xspec} and are reported in Supplementary Table~2. 
We verified the consistency of the spectro-polarimetric and binned analysis results for all bins and found no significant differences in the phase dependence of the PD and PA, therefore only the results of the spectro-polarimetric analysis are reported.

The same procedure has been used to investigate the time dependence of the polarization properties over the observation. 
The full dataset was split into seven intervals separated by large gaps defined either by the instrumental good time intervals or by the eclipses of the source. 
For each interval, the Stokes spectra ($I/Q/U)$ were extracted and jointly modelled using \texttt{nthcomp} and \texttt{polconst} models to estimate the PD and PA values. 
The value of the power-law index in \texttt{nthcomp} model was considered as a free parameter to accommodate possible minor changes in the spectral shape over the observation. 
The final values and uncertainties were estimated based on \textit{mcmc} chains produced using \texttt{chain} command in \textsc{xspec} and are reported in Supplementary Table~3. 
Again, we verified consistency of the spectro-polarimetric and the binned analysis results for all bins and found no significant differences in the phase dependence of the PD and PA, therefore again only results of the spectro-polarimetric analysis are reported.

\subsection*{Modelling polarization from heated neutron star atmosphere}

Polarization from a strongly magnetized accreting neutron star is largely defined by the emission region structure which is not known. 
Earlier estimates for Her~X-1 \cite{2021MNRAS.501..129C} were based on the accretion column model \cite{2016ApJ...831..194W} which seems to be consistent with the observed broadband spectrum.
The observed polarization, however, is significantly lower ($\sim$5--15\%) than the predicted one (60--80\%), requiring modifications to the models.
There are several mechanisms that may depolarize radiation as it leaves the accretion column and travels through the magnetosphere. 
For instance, the depolarization can be caused by passing of radiation from the accretion column through the so-called vacuum resonance, where the contributions of plasma and magnetized vacuum to the dielectric tensor cancel each other and fast transformation of the normal modes of radiation occurs \cite{1978SvAL....4..117G,1979JETP...49..741P}. 
If the place where the final scattering of  radiation takes place (i.e. the photosphere) also lies in this region, we expect substantial Faraday depolarization reducing the PD without changing the spectral energy distribution or the pulse profile. 
Furthermore, as the radiation travels from the column through the magnetosphere, generally it will pass through a region where the direction of propagation is nearly parallel to the magnetic field lines.
Depending on the geometry of the emission region and the photon energy this can also result in substantial depolarization \cite{2021MNRAS.501..109C}.

On the other hand, it is unclear whether an accretion column is present at all in Her~X-1. 
Although the observed luminosity is close to the critical value \cite{2015MNRAS.447.1847M}, the source demonstrates a positive correlation of the cyclotron line energy with luminosity \cite{2007A&A...465L..25S}, which implies that the accreting pulsar is in a sub-critical state, when the energy of the infalling matter is dissipated at the neutron star surface but not in a radiation-dominated shock above it. 
In such a situation, fast ions of the accretion flow heat the neutron star atmosphere, and the thermal photons emerging from this heated atmosphere back-scatter on the in-falling electrons of the accretion flow with a corresponding energy gain (bulk Comptonization), and these back-scattered photons additionally heat the upper atmosphere. 
If the local mass accretion rate  is close to the critical one, almost all the emergent photons will be back-scattered, and, as a result, radiation  escapes primarily along the tangential direction to the neutron star surface, forming a ``fan''-like angular distribution of the escaping radiation helping to explain the observed high pulsed fraction.   
An accurate self-consistent numerical model describing the processes above is yet to be developed.  
Here we consider a toy model of the overheated magnetized model atmosphere to demonstrate how the observed low polarization can be produced.  
Such models have been used for interpretation of accreting neutron stars \cite{1969SvA....13..175Z,2018A&A...619A.114S,2019MNRAS.483..599G,2021MNRAS.503.5193M} although it is important to emphasise that the broadband spectrum of Her~X-1 is clearly not described by any of these models alone.

In this simplified picture, the key process that is responsible for low polarization is a mode conversion at the vacuum resonance. 
For a given photon energy and magnetic field strength, the vacuum resonance occurs at a plasma density \cite{1979JETP...49..741P} of $\rho_{\rm V} \approx 10^{-4} (B/10^{12}~\mbox{G})^2 E^2_{\rm keV}$\,g\,cm$^{-3}$ . 
At that density, the contribution of the virtual electron-positron pairs to the dielectric tensor becomes equal to the plasma contribution, and the ordinary (O) and extraordinary (X) modes of radiation can convert to each other. 
Here we consider the radiation transfer in magnetized plasma in the approximation of these two modes instead of the full description in terms of Stokes parameters. 
We found that the modes become close to each other at a given photon energy in the emergent spectrum if the vacuum resonance is located in the transition atmospheric layer with a strong temperature gradient from the upper overheated layer of a temperature a few tens of keVs to the lower layer of the atmosphere where the temperature is about 1--2 keV.

We illustrate this statement with a toy model of the transition region between two atmospheric parts  (see Supplementary Figure 2A). 
We assume the surface magnetic field strength $B = 4\times 10^{12}$\,G, the  temperature of the overheated layers $T_{\rm up}= 3.1 \times 10^8$\,K, and the temperature of the bottom cold atmosphere $T_{\rm low} = 1.5 \times 10^7$\,K.
We consider three different transition depths of $m_{\rm up} =$\,0.3, 3, and 30~g~cm$^{-2}$.  
The corresponding gas pressure is determined by the product of the column density of  plasma $m$ and the surface gravity $g$, $P_{\rm gas}= gm$, computed using the neutron star mass $M = 1.4 M_{\odot}$ and radius $R = 12$\,km. 
For the temperature structure we adopt the dependence 
\begin{equation} 
T(m) = \frac{T_{\rm up}-T_{\rm low}}{\exp[6(m/m_{\rm up}-1)]+1} + T_{\rm low}.
\end{equation}
 
We solved the radiation transfer equation for the two modes using the magnetic opacities and the mode conversion as described in \cite{2006MNRAS.373.1495V}, with no external radiation flux as the upper boundary condition and the Planck function for the intensity as the lower boundary condition \cite{2009A&A...500..891S}. 
The polarization fraction of the emergent flux in the observed energy band with and without mode conversion is shown in Supplementary Figure 2B. 
The model with the transition depth $m_{\rm up} = 3$\,g\,cm$^{-2}$ demonstrates a low polarization, which is explained by the mode conversion at the transition region with the strong temperature gradient, see Supplementary Figure~3. 
We note that models with either thinner or thicker overheated layers yield a higher polarization degree (i.e. a larger fraction of total flux is in one of the modes); however, the dominant modes are different in these cases (Supplementary Figure 2B). 
If the thickness of the upper layer is low, $m_{\rm up}$=\,0.3\,g\,cm$^{-2}$, the vacuum resonance occurs in the cold inner part of the atmosphere with strong mode conversion. 
As a result, the O-mode dominates. 
On the other hand, the mode conversion is inefficient if the vacuum resonance occurs within the overheated layer with $m_{\rm up}$=\,30\,g\,cm$^{-2}$, so the X-mode dominates.  
Note that the depth of the transition layer of $m_{\rm up} \approx 3$\,g\,cm$^{-2}$ appears to be natural as it corresponds to the optical depth of around unity, where the free-free cooling becomes inefficient while the Compton cooling becomes important.
The radiation escaping the atmosphere can be dominated by the O- or X-mode, depending on the exact value of $m_{\rm up}$ and the detailed temperature structure. 
The polarization mode can also depend on the angle between the surface normal and the direction of photon propagation. 
At energies a factor of 10 below the electron cyclotron energy, the vacuum polarization dominates at the outer overheated layer.
As a result, both modes are nearly linearly polarized at zenith angles larger than $\sim6\degr$ and therefore in a broad angle range the PD can be computed as the ratio of the difference in the intensities of the two modes to their sum \cite{2000ApJ...529.1011P}. 
As an illustration, we show in  Supplementary Figure 2C the PD as observed at different zenith angles for $m_{\rm up}=3$\,g\,cm$^{-2}$. 
We see that at very small and very large inclination the PD is negative (i.e. the X-mode dominates), while at intermediate angles the PD is positive (i.e. the O-mode dominates).
This indicates that mixing of radiation observed from different emission regions (i.e. seen at different zenith angles) can lead to depolarization.
We cannot confidently state that the suggested process is responsible for the low polarization of the observed radiation from Her X-1, but it can be potentially important for the final accurate model and for the interpretation of the low polarization signal from other X-ray sources, e.g., magnetars.

\section{Data Availability}
IXPE data and analysis tools are publicly available from HEASARC data archive (\url{https://heasarc.gsfc.nasa.gov}). Optical polarimetry data used in the paper are published in \cite{1991A&A...244L..41E}. 

\section{Acknowledgements}
This paper is based on the observations made by the Imaging X-ray Polarimetry Explorer (IXPE), a joint US and Italian mission.  This research used data products provided by the IXPE Team (MSFC, SSDC, INAF, and INFN) and distributed with additional software tools by the High-Energy Astrophysics Science Archive Research Center (HEASARC), at NASA Goddard Space Flight Center (GSFC).
The US contribution is supported by the National Aeronautics and Space Administration (NASA) and led and managed by its Marshall Space Flight Center (MSFC), with industry partner Ball Aerospace (contract NNM15AA18C). 
The Italian contribution is supported by the Italian Space Agency (Agenzia Spaziale Italiana, ASI) through contract ASI-OHBI-2017-12-I.0, agreements ASI-INAF-2017-12-H0 and ASI-INFN-2017.13-H0, and its Space Science Data Center (SSDC), and by the Istituto Nazionale di Astrofisica (INAF) and the Istituto Nazionale di Fisica Nucleare (INFN) in Italy.
V.D. and V.F.S. acknowledge support from the German Academic Exchange Service (DAAD) travel grant 57525212. 
J.P. and  S.S.T. thank the Russian Science Foundation grant 20-12-00364 and the Academy of Finland grants 333112, 349144, 349373, and 349906 for support.
V.F.S. thanks the German Research Foundation (DFG) grant WE 1312/53-1. 
I.C. is a Sherman Fairchild Fellow at Caltech and thanks the Burke Institute at Caltech for supporting her research.
A.A.M. acknowledges support from the Netherlands Organization for Scientific Research Veni Fellowship.

\section{Author Contributions Statement}
V.D. analysed the data and wrote the draft of the manuscript. J.P. led the work of the IXPE Topical Working Group on Accreting Neutron Stars and contributed to modelling geometrical parameters, interpretation and the text.  S.S.T. produced an independent analysis of the data. V.F.S. led modelling of the polarization from heated atmospheres. A.D.M., F.L.M., F.M., and J.R. provided quick-look analysis of the data and energy scale correction calculation. I.C., J.H., A.A.M., S.Z., R.S. and A.S. contributed to interpretation of the results and writing of the text. M.B. and G.G.P. acted as internal referees of the paper and contributed to interpretation. Other members of the IXPE collaboration contributed to the design of the mission and its science case and planning of the observations. All authors provided input and comments on the manuscript.
\section{Competing Interests Statement}
The authors declare no competing interests. 

\clearpage 

\section{Tables}
\begin{table}[h!]
 \centering
\caption{\textbf{Orbital and pulsar geometrical parameters of Her~X-1.}
}          
\centering          
\begin{tabular}{ccccc}     
\hline\hline  
$\chi_{\rm p,*}$  & $\theta$ & $i_{\rm p}$ & 
$\chi_{\rm orb,*}$  & $i_{\rm orb}$  \\ 
deg & deg & deg & deg & deg  \\ 
\hline 
$56.9\pm1.6$  & $12.1\pm3.7$ & Eq.~(\ref{eq:distr_inclp}) & $28.9\pm5.9$ & $100.4\pm4.9$ \\
\hline                  
\end{tabular}
\label{tab:geometry}      
\end{table} 

\clearpage
\newpage
\section{Figure Legends/Captions}

\renewcommand\thefigure{Extended Data Figure \arabic{figure}}  
\renewcommand{\figurename}{}
\setcounter{figure}{0}
\begin{figure}[h!]
\centering\includegraphics[width=\columnwidth]{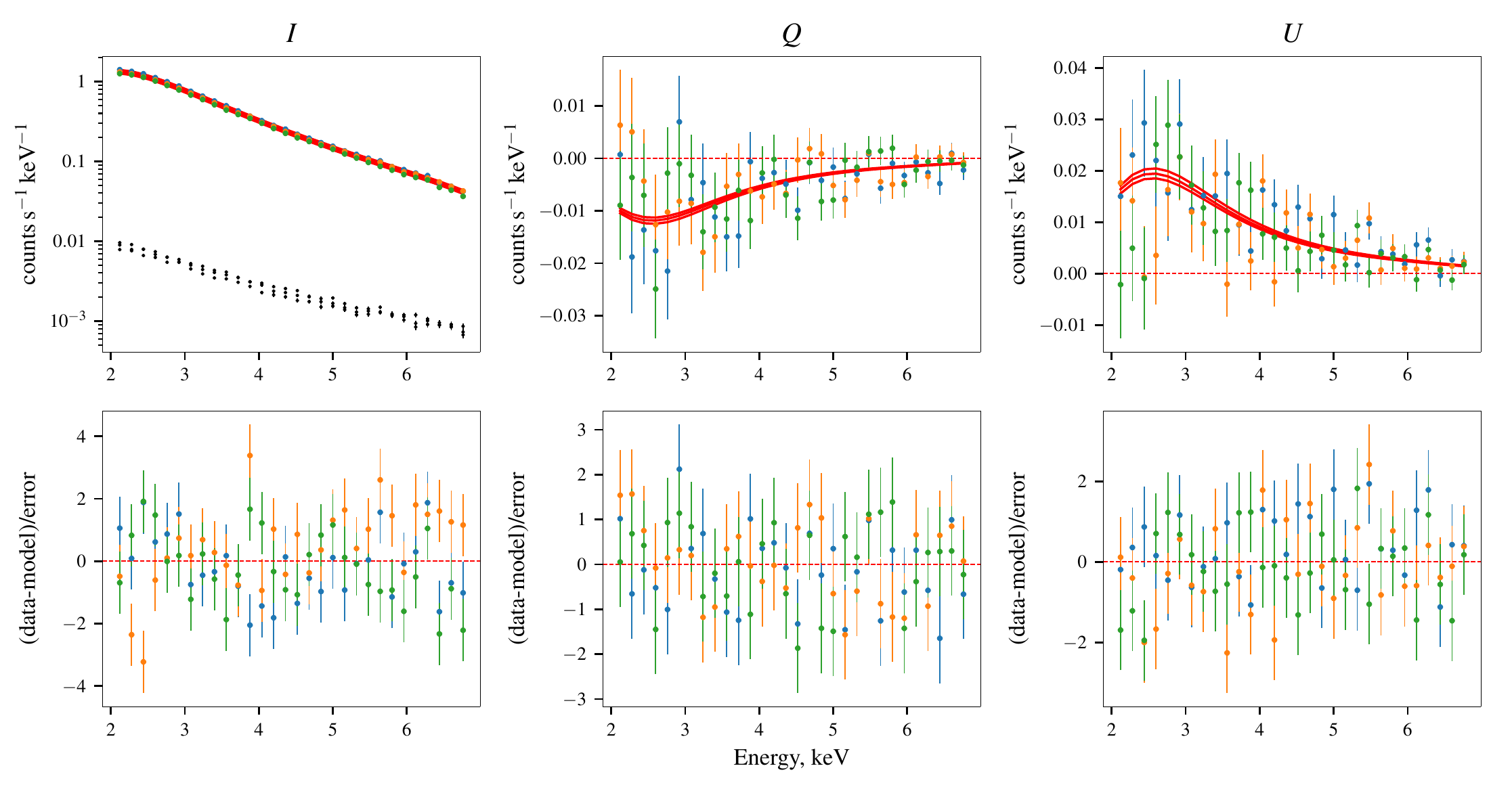}
\caption{\textbf{Observed Stokes spectra of Her~X-1.} 
The top row shows spectra of the three Stokes parameters $I$, $Q$, and $U$, while the bottom row shows the residuals to the best-fitting model (\texttt{nthcomp} for intensity and \texttt{polconst} for $Q$ and $U$). The results for the three detector units are colour-coded, the black points in the first panel show the estimated background level for each detector. }
\label{fig:spefit}
\end{figure}

\begin{figure}[h!]
\centering\includegraphics[width=0.90\textwidth]{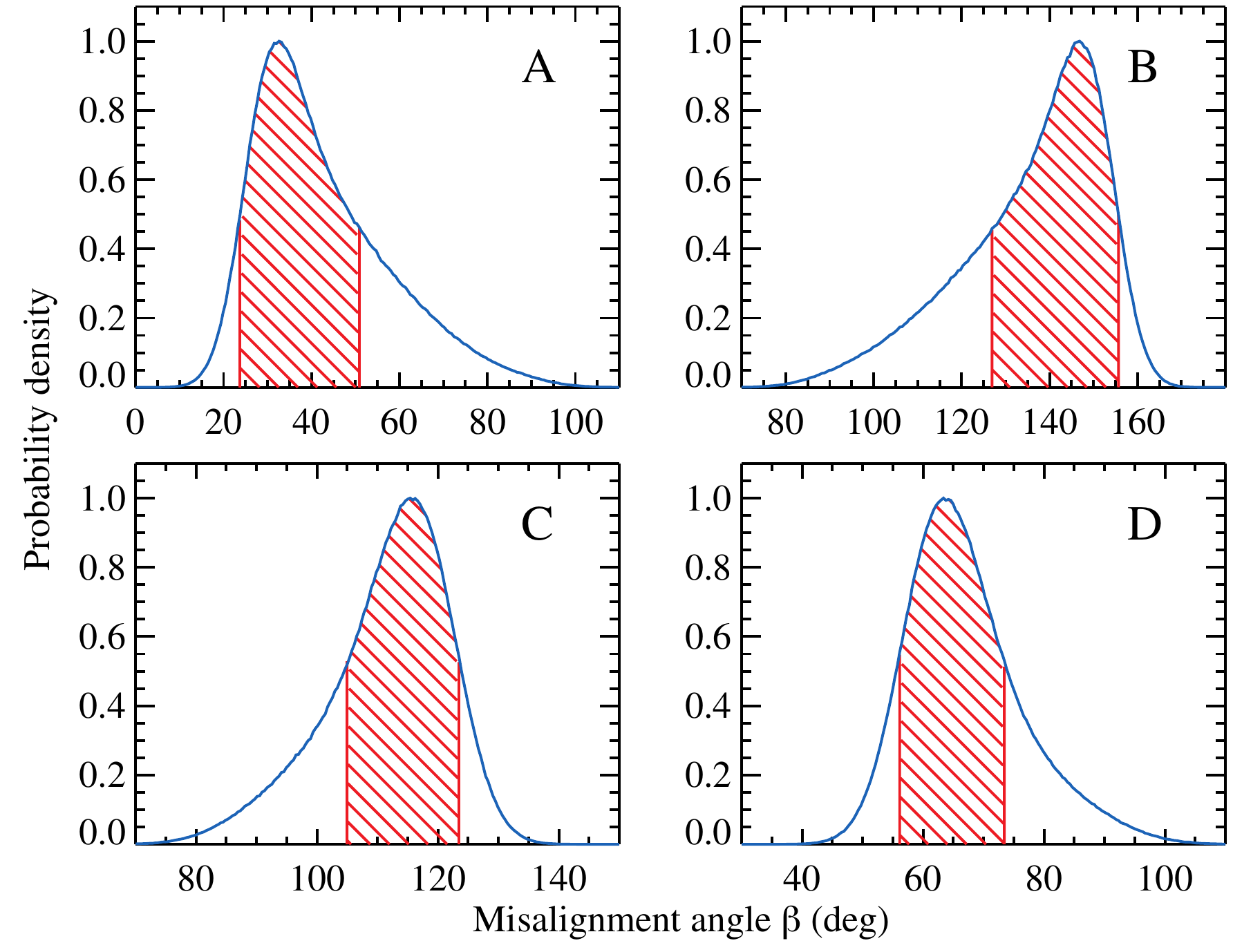}
\caption{\textbf{Probability distribution function for the misalignment angle.} 
The distribution normalized to the peak value is shown for the misalignment angle between the pulsar and the orbital angular momenta. 
The red hatched region corresponds to the 68\% confidence interval (i.e. between 16th and 84th percentiles of the posterior probability distribution). 
Four panels correspond to four different cases for the choice of $\chi_{\rm p}$:
\textbf{(A)} $\chi_{\rm p}=\chi_{\rm p,*}=56\fdg9\pm1\fdg6$; 
\textbf{(B)} $\chi_{\rm p}=\chi_{\rm p,*}+180\degr$;
\textbf{(C)} $\chi_{\rm p}=\chi_{\rm p,*}+90\degr$; 
\textbf{(D)} $\chi_{\rm p}=\chi_{\rm p,*}-90\degr$. Here we take $\chi_{\rm orb}=\chi_{\rm orb,*}=28\fdg9\pm5\fdg9$. 
\label{fig:mis_angle}} 
\end{figure}

\clearpage
\newpage
\section{References}
\renewcommand\refname{}


\begin{thebibliography}{10}
\expandafter\ifx\csname url\endcsname\relax
  \def\url#1{\burl{#1}}\fi
\expandafter\ifx\csname urlprefix\endcsname\relax\def\urlprefix{URL }\fi
\providecommand{\bibinfo}[2]{#2}
\providecommand{\eprint}[2][]{\url{#2}}
\providecommand{\doi}[1]{\url{https://doi.org/#1}}
\bibcommenthead

\bibitem{1972ApJ...174L.143T}
\bibinfo{author}{{Tananbaum}, H.} \emph{et~al.}
\newblock \bibinfo{title}{{Discovery of a Periodic Pulsating Binary X-Ray
  Source in Hercules from UHURU}}.
\newblock \emph{\bibinfo{journal}{\apjl}} \textbf{\bibinfo{volume}{174}},
  \bibinfo{pages}{L143} (\bibinfo{year}{1972}).

\bibitem{bj21}
\bibinfo{author}{{Bailer-Jones}, C.~A.~L.}, \bibinfo{author}{{Rybizki}, J.},
  \bibinfo{author}{{Fouesneau}, M.}, \bibinfo{author}{{Demleitner}, M.} \&
  \bibinfo{author}{{Andrae}, R.}
\newblock \bibinfo{title}{{Estimating Distances from Parallaxes. V. Geometric
  and Photogeometric Distances to 1.47 Billion Stars in Gaia Early Data Release
  3}}.
\newblock \emph{\bibinfo{journal}{\aj}} \textbf{\bibinfo{volume}{161}},
  \bibinfo{pages}{147} (\bibinfo{year}{2021}).

\bibitem{1976ApJ...208..567M}
\bibinfo{author}{{Middleditch}, J.} \& \bibinfo{author}{{Nelson}, J.}
\newblock \bibinfo{title}{{Studies of optical pulsations from HZ
  Herculis/Hercules X-1: a determination of the mass of the neutron star.}}
\newblock \emph{\bibinfo{journal}{\apj}} \textbf{\bibinfo{volume}{208}},
  \bibinfo{pages}{567--586} (\bibinfo{year}{1976}).

\bibitem{1978ApJ...219L.105T}
\bibinfo{author}{{Truemper}, J.} \emph{et~al.}
\newblock \bibinfo{title}{{Evidence for strong cyclotron line emission in the
  hard X-ray spectrum of Hercules X-1.}}
\newblock \emph{\bibinfo{journal}{\apjl}} \textbf{\bibinfo{volume}{219}},
  \bibinfo{pages}{L105--L110} (\bibinfo{year}{1978}).

\bibitem{1973ApJ...184..227G}
\bibinfo{author}{{Giacconi}, R.} \emph{et~al.}
\newblock \bibinfo{title}{{Further X-ray observations of Hercules X-1 from
  Uhuru.}}
\newblock \emph{\bibinfo{journal}{\apj}} \textbf{\bibinfo{volume}{184}},
  \bibinfo{pages}{227} (\bibinfo{year}{1973}).

\bibitem{1986ApJ...300L..63T}
\bibinfo{author}{{Truemper}, J.}, \bibinfo{author}{{Kahabka}, P.},
  \bibinfo{author}{{Oegelman}, H.}, \bibinfo{author}{{Pietsch}, W.} \&
  \bibinfo{author}{{Voges}, W.}
\newblock \bibinfo{title}{{EXOSAT Observations of the 35 Day Cycle of Hercules
  X-1: Evidence for Neutron Star Precession}}.
\newblock \emph{\bibinfo{journal}{\apjl}} \textbf{\bibinfo{volume}{300}},
  \bibinfo{pages}{L63} (\bibinfo{year}{1986}).

\bibitem{2009A&A...494.1025S}
\bibinfo{author}{{Staubert}, R.} \emph{et~al.}
\newblock \bibinfo{title}{{Two \raisebox{-0.5ex}\textasciitilde35 day clocks in
  Hercules X-1: evidence for neutron star free precession}}.
\newblock \emph{\bibinfo{journal}{\aap}} \textbf{\bibinfo{volume}{494}},
  \bibinfo{pages}{1025--1030} (\bibinfo{year}{2009}).

\bibitem{2013MNRAS.435.1147P}
\bibinfo{author}{{Postnov}, K.} \emph{et~al.}
\newblock \bibinfo{title}{{Variable neutron star free precession in Hercules
  X-1 from evolution of RXTE X-ray pulse profiles with phase of the 35-d
  cycle}}.
\newblock \emph{\bibinfo{journal}{\mnras}} \textbf{\bibinfo{volume}{435}},
  \bibinfo{pages}{1147--1164} (\bibinfo{year}{2013}).

\bibitem{2021MNRAS.501..129C}
\bibinfo{author}{{Caiazzo}, I.} \& \bibinfo{author}{{Heyl}, J.}
\newblock \bibinfo{title}{{Polarization of accreting X-ray pulsars - II.
  Hercules X-1}}.
\newblock \emph{\bibinfo{journal}{\mnras}} \textbf{\bibinfo{volume}{501}},
  \bibinfo{pages}{129--136} (\bibinfo{year}{2021}).

\bibitem{2003MNRAS.342..446L}
\bibinfo{author}{{Leahy}, D.~A.}
\newblock \bibinfo{title}{{Modelling the extreme ultraviolet emission during
  the low state of Hercules X-1}}.
\newblock \emph{\bibinfo{journal}{\mnras}} \textbf{\bibinfo{volume}{342}},
  \bibinfo{pages}{446--452} (\bibinfo{year}{2003}).

\bibitem{2006AstL...32..804K}
\bibinfo{author}{{Klochkov}, D.~K.} \emph{et~al.}
\newblock \bibinfo{title}{{Observational manifestations of the change in the
  tilt of the accretion disc to the orbital plane in Her X-1/HZ Her with phase
  of its 35-day period}}.
\newblock \emph{\bibinfo{journal}{Astronomy Letters}}
  \textbf{\bibinfo{volume}{32}}, \bibinfo{pages}{804--815}
  (\bibinfo{year}{2006}).

\bibitem{2021Univ....7..160L}
\bibinfo{author}{{Leahy}, D.} \& \bibinfo{author}{{Wang}, Y.}
\newblock \bibinfo{title}{{The 35-Day Cycle of Hercules X-1 in Multiple Energy
  Bands from MAXI and Swift/BAT Monitoring}}.
\newblock \emph{\bibinfo{journal}{Universe}} \textbf{\bibinfo{volume}{7}},
  \bibinfo{pages}{160} (\bibinfo{year}{2021}).

\bibitem{2012MNRAS.425....8I}
\bibinfo{author}{{Igna}, C.~D.} \& \bibinfo{author}{{Leahy}, D.~A.}
\newblock \bibinfo{title}{{Light-curve dip production through accretion
  stream-accretion disc impact in the HZ Her/Her X-1 binary star system}}.
\newblock \emph{\bibinfo{journal}{\mnras}} \textbf{\bibinfo{volume}{425}},
  \bibinfo{pages}{8--20} (\bibinfo{year}{2012}).

\bibitem{1999A&A...348..917S}
\bibinfo{author}{{Shakura}, N.~I.}, \bibinfo{author}{{Prokhorov}, M.~E.},
  \bibinfo{author}{{Postnov}, K.~A.} \& \bibinfo{author}{{Ketsaris}, N.~A.}
\newblock \bibinfo{title}{{On the origin of X-ray dips in Her X-1}}.
\newblock \emph{\bibinfo{journal}{\aap}} \textbf{\bibinfo{volume}{348}},
  \bibinfo{pages}{917--923} (\bibinfo{year}{1999}).

\bibitem{2015APh....64...40K}
\bibinfo{author}{{Kislat}, F.}, \bibinfo{author}{{Beilicke}, M.},
  \bibinfo{author}{{Guo}, Q.}, \bibinfo{author}{{Zajczyk}, A.} \&
  \bibinfo{author}{{Krawczynski}, H.}
\newblock \bibinfo{title}{{An unfolding method for X-ray spectro-polarimetry}}.
\newblock \emph{\bibinfo{journal}{Astroparticle Physics}}
  \textbf{\bibinfo{volume}{64}}, \bibinfo{pages}{40--48}
  (\bibinfo{year}{2015}).

\bibitem{2017ApJ...838...72S}
\bibinfo{author}{{Strohmayer}, T.~E.}
\newblock \bibinfo{title}{{X-ray Spectro-polarimetry with Photoelectric Polarimeters}}.
\newblock \emph{\bibinfo{journal}{\apj}} \textbf{\bibinfo{volume}{838}},
  \bibinfo{pages}{72} (\bibinfo{year}{2017}).

\bibitem{1978SvAL....4..117G}
\bibinfo{author}{{Gnedin}, Y.~N.}, \bibinfo{author}{{Pavlov}, G.~G.} \&
  \bibinfo{author}{{Shibanov}, Y.~A.}
\newblock \bibinfo{title}{{The effect of vacuum birefringence in a magnetic
  field on the polarization and beaming of X-ray pulsars}}.
\newblock \emph{\bibinfo{journal}{Soviet Astronomy Letters}}
  \textbf{\bibinfo{volume}{4}}, \bibinfo{pages}{117--119}
  (\bibinfo{year}{1978}).

\bibitem{1979JETP...49..741P}
\bibinfo{author}{{Pavlov}, G.~G.} \& \bibinfo{author}{{Shibanov}, Y.~A.}
\newblock \bibinfo{title}{{Influence of vacuum polarization by a magnetic field
  on the propagation of electromagnetic waves in a plasma}}.
\newblock \emph{\bibinfo{journal}{Soviet Journal of Experimental and
  Theoretical Physics}} \textbf{\bibinfo{volume}{49}}, \bibinfo{pages}{741}
  (\bibinfo{year}{1979}).

\bibitem{heylsh00}
\bibinfo{author}{{Heyl}, J.~S.} \& \bibinfo{author}{{Shaviv}, N.~J.}
\newblock \bibinfo{title}{{Polarization evolution in strong magnetic fields}}.
\newblock \emph{\bibinfo{journal}{\mnras}} \textbf{\bibinfo{volume}{311}},
  \bibinfo{pages}{555--564} (\bibinfo{year}{2000}).

\bibitem{2018Galax...6...76H}
\bibinfo{author}{{Heyl}, J.} \& \bibinfo{author}{{Caiazzo}, I.}
\newblock \bibinfo{title}{{Strongly Magnetized Sources: QED and X-ray
  Polarization}}.
\newblock \emph{\bibinfo{journal}{Galaxies}} \textbf{\bibinfo{volume}{6}},
  \bibinfo{pages}{76} (\bibinfo{year}{2018}).

\bibitem{2000ApJ...529..968B}
\bibinfo{author}{{Blum}, S.} \& \bibinfo{author}{{Kraus}, U.}
\newblock \bibinfo{title}{{Analyzing X-Ray Pulsar Profiles: Geometry and Beam
  Pattern of Hercules X-1}}.
\newblock \emph{\bibinfo{journal}{\apj}} \textbf{\bibinfo{volume}{529}},
  \bibinfo{pages}{968--977} (\bibinfo{year}{2000}).

\bibitem{RC69}
\bibinfo{author}{{Radhakrishnan}, V.} \& \bibinfo{author}{{Cooke}, D.~J.}
\newblock \bibinfo{title}{{Magnetic Poles and the Polarization Structure of
  Pulsar Radiation}}.
\newblock \emph{\bibinfo{journal}{\aplett}} \textbf{\bibinfo{volume}{3}},
  \bibinfo{pages}{225} (\bibinfo{year}{1969}).

\bibitem{Drissen86}
\bibinfo{author}{{Drissen}, L.}, \bibinfo{author}{{Lamontagne}, R.},
  \bibinfo{author}{{Moffat}, A.~F.~J.}, \bibinfo{author}{{Bastien}, P.} \&
  \bibinfo{author}{{Seguin}, M.}
\newblock \bibinfo{title}{{Spectroscopic and Polarimetric Parameters of the
  Runaway WN7 Binary System HD 197406: Is the Secondary an X-ray--quiet Black
  Hole?}}
\newblock \emph{\bibinfo{journal}{\apj}} \textbf{\bibinfo{volume}{304}},
  \bibinfo{pages}{188} (\bibinfo{year}{1986}).

\bibitem{2022MNRAS.513.3359K}
\bibinfo{author}{{Kolesnikov}, D.}, \bibinfo{author}{{Shakura}, N.} \&
  \bibinfo{author}{{Postnov}, K.}
\newblock \bibinfo{title}{{Evidence for neutron star triaxial free precession
  in Her X-1 from Fermi/GBM pulse period measurements}}.
\newblock \emph{\bibinfo{journal}{\mnras}} \textbf{\bibinfo{volume}{513}},
  \bibinfo{pages}{3359--3367} (\bibinfo{year}{2022}).

\bibitem{1991A&A...244L..41E}
\bibinfo{author}{{Egonsson}, J.} \& \bibinfo{author}{{Hakala}, P.}
\newblock \bibinfo{title}{{Discovery of variable optical polarization in Her
  X-1.}}
\newblock \emph{\bibinfo{journal}{\aap}} \textbf{\bibinfo{volume}{244}},
  \bibinfo{pages}{L41--L42} (\bibinfo{year}{1991}).

\bibitem{2021A&A...648A..39S}
\bibinfo{author}{{Shakura}, N.~I.} \emph{et~al.}
\newblock \bibinfo{title}{{Observations of Her X-1 in low states during
  SRG/eROSITA all-sky survey}}.
\newblock \emph{\bibinfo{journal}{\aap}} \textbf{\bibinfo{volume}{648}},
  \bibinfo{pages}{A39} (\bibinfo{year}{2021}).

\bibitem{1991ApJ...378..696P}
\bibinfo{author}{{Petterson}, J.~A.}, \bibinfo{author}{{Rothschild}, R.~E.} \&
  \bibinfo{author}{{Gruber}, D.~E.}
\newblock \bibinfo{title}{{A Model for the 35 Day Variations in the Pulse
  Profile of Hercules X-1}}.
\newblock \emph{\bibinfo{journal}{\apj}} \textbf{\bibinfo{volume}{378}},
  \bibinfo{pages}{696} (\bibinfo{year}{1991}).

\bibitem{Biryukov21}
\bibinfo{author}{{Biryukov}, A.} \& \bibinfo{author}{{Abolmasov}, P.}
\newblock \bibinfo{title}{{Magnetic angle evolution in accreting neutron
  stars}}.
\newblock \emph{\bibinfo{journal}{\mnras}} \textbf{\bibinfo{volume}{505}},
  \bibinfo{pages}{1775--1786} (\bibinfo{year}{2021}).

\bibitem{2021MNRAS.501..109C}
\bibinfo{author}{{Caiazzo}, I.} \& \bibinfo{author}{{Heyl}, J.}
\newblock \bibinfo{title}{{Polarization of accreting X-ray pulsars. I. A new
  model}}.
\newblock \emph{\bibinfo{journal}{\mnras}} \textbf{\bibinfo{volume}{501}},
  \bibinfo{pages}{109--128} (\bibinfo{year}{2021}).

\bibitem{2022MNRAS.515..571M}
\bibinfo{author}{{M{\"o}nkk{\"o}nen}, J.} \emph{et~al.}
\newblock \bibinfo{title}{{Constraints on the magnetic field structure in
  accreting compact objects from aperiodic variability}}.
\newblock \emph{\bibinfo{journal}{\mnras}} \textbf{\bibinfo{volume}{515}},
  \bibinfo{pages}{571--580} (\bibinfo{year}{2022}).

\bibitem{Weisskopf2022}
\bibinfo{author}{{Weisskopf}, M.~C.} \emph{et~al.}
\newblock \bibinfo{title}{{The Imaging X-Ray Polarimetry Explorer (IXPE):
  Pre-Launch}}.
\newblock \emph{\bibinfo{journal}{J. Astron. Telesc. Instrum. Syst.}}
  \textbf{\bibinfo{volume}{8}}, \bibinfo{pages}{026002} (\bibinfo{year}{2022}).

\bibitem{Soffitta2021}
\bibinfo{author}{{Soffitta}, P.} \emph{et~al.}
\newblock \bibinfo{title}{{The Instrument of the Imaging X-Ray Polarimetry
  Explorer}}.
\newblock \emph{\bibinfo{journal}{\aj}} \textbf{\bibinfo{volume}{162}},
  \bibinfo{pages}{208} (\bibinfo{year}{2021}).

\bibitem{Baldini2021}
\bibinfo{author}{{Baldini}, L.} \emph{et~al.}
\newblock \bibinfo{title}{{Design, construction, and test of the Gas Pixel
  Detectors for the IXPE mission}}.
\newblock \emph{\bibinfo{journal}{Astroparticle Physics}}
  \textbf{\bibinfo{volume}{133}}, \bibinfo{pages}{102628}
  (\bibinfo{year}{2021}).

\bibitem{2015APh....68...45K}
\bibinfo{author}{{Kislat}, F.}, \bibinfo{author}{{Clark}, B.},
  \bibinfo{author}{{Beilicke}, M.} \& \bibinfo{author}{{Krawczynski}, H.}
\newblock \bibinfo{title}{{Analyzing the data from X-ray polarimeters with
  Stokes parameters}}.
\newblock \emph{\bibinfo{journal}{Astroparticle Physics}}
  \textbf{\bibinfo{volume}{68}}, \bibinfo{pages}{45--51}
  (\bibinfo{year}{2015}).

\bibitem{Baldini2022}
\bibinfo{author}{{Baldini}, L.} \emph{et~al.}
\newblock \bibinfo{title}{{ixpeobssim: a Simulation and Analysis Framework for
  the Imaging X-ray Polarimetry Explorer}}.
\newblock \emph{\bibinfo{journal}{arXiv e-prints}}
  \bibinfo{pages}{arXiv:2203.06384} (\bibinfo{year}{2022}).

\bibitem{Arnaud1996}
\bibinfo{author}{{Arnaud}, K.~A.}
\newblock \bibinfo{title}{{XSPEC: The First Ten Years}}.
\newblock In \bibinfo{editor}{{Jacoby}, G.~H.} \& \bibinfo{editor}{{Barnes},
  J.} (eds.) \emph{\bibinfo{booktitle}{Astronomical Data Analysis Software and
  Systems V}}, vol. \bibinfo{volume}{101} of
  \emph{\bibinfo{series}{Astronomical Society of the Pacific Conference
  Series}}, \bibinfo{pages}{17} (\bibinfo{year}{1996}).

\bibitem{Rankin_2022}
\bibinfo{author}{{Rankin}, J.} \emph{et~al.}
\newblock \bibinfo{title}{{An Algorithm to Calibrate and Correct the Response
  to Unpolarized Radiation of the X-Ray Polarimeter Onboard IXPE}}.
\newblock \emph{\bibinfo{journal}{\aj}} \textbf{\bibinfo{volume}{163}},
  \bibinfo{pages}{39} (\bibinfo{year}{2022}).

\bibitem{Di_Marco_2022}
\bibinfo{author}{{Di Marco}, A.} \emph{et~al.}
\newblock \bibinfo{title}{{A Weighted Analysis to Improve the X-Ray
  Polarization Sensitivity of the Imaging X-ray Polarimetry Explorer}}.
\newblock \emph{\bibinfo{journal}{\aj}} \textbf{\bibinfo{volume}{163}},
  \bibinfo{pages}{170} (\bibinfo{year}{2022}).

\bibitem{1999MNRAS.309..561Z}
\bibinfo{author}{{{\.Z}ycki}, P.~T.}, \bibinfo{author}{{Done}, C.} \&
  \bibinfo{author}{{Smith}, D.~A.}
\newblock \bibinfo{title}{{The 1989 May outburst of the soft X-ray transient GS
  2023+338 (V404 Cyg)}}.
\newblock \emph{\bibinfo{journal}{\mnras}} \textbf{\bibinfo{volume}{309}},
  \bibinfo{pages}{561--575} (\bibinfo{year}{1999}).

\bibitem{bic_ref}
\bibinfo{author}{Wit, E.}, \bibinfo{author}{{van den Heuvel}, E.} \&
  \bibinfo{author}{Romeijn, J.-W.}
\newblock \bibinfo{title}{All models are wrong...': an introduction to model
  uncertainty}.
\newblock \emph{\bibinfo{journal}{Statistica Neerlandica}}
  \textbf{\bibinfo{volume}{66}}, \bibinfo{pages}{217--236}
  (\bibinfo{year}{2012}).

\bibitem{2009A&A...500..883S}
\bibinfo{author}{{Staubert}, R.}, \bibinfo{author}{{Klochkov}, D.} \&
  \bibinfo{author}{{Wilms}, J.}
\newblock \bibinfo{title}{{Updating the orbital ephemeris of Hercules X-1; rate
  of decay and eccentricity of the orbit}}.
\newblock \emph{\bibinfo{journal}{\aap}} \textbf{\bibinfo{volume}{500}},
  \bibinfo{pages}{883--889} (\bibinfo{year}{2009}).

\bibitem{Lomb76}
\bibinfo{author}{{Lomb}, N.~R.}
\newblock \bibinfo{title}{{Least-Squares Frequency Analysis of Unequally Spaced
  Data}}.
\newblock \emph{\bibinfo{journal}{\apss}} \textbf{\bibinfo{volume}{39}},
  \bibinfo{pages}{447--462} (\bibinfo{year}{1976}).

\bibitem{Scargle82}
\bibinfo{author}{{Scargle}, J.~D.}
\newblock \bibinfo{title}{{Studies in astronomical time series analysis. II.
  Statistical aspects of spectral analysis of unevenly spaced data.}}
\newblock \emph{\bibinfo{journal}{\apj}} \textbf{\bibinfo{volume}{263}},
  \bibinfo{pages}{835--853} (\bibinfo{year}{1982}).

\bibitem{2001ESASP.459..309K}
\bibinfo{author}{{Kuster}, M.} \emph{et~al.}
\newblock \bibinfo{title}{{Evolution of the 1.24s pulse profile during a Her
  X-1 turn-on}}.
\newblock In \bibinfo{editor}{{Gimenez}, A.}, \bibinfo{editor}{{Reglero}, V.}
  \& \bibinfo{editor}{{Winkler}, C.} (eds.) \emph{\bibinfo{booktitle}{Exploring
  the Gamma-Ray Universe}}, vol. \bibinfo{volume}{459} of
  \emph{\bibinfo{series}{ESA Special Publication}}, \bibinfo{pages}{309--312}
  (\bibinfo{year}{2001}).

\bibitem{Poutanen20RVM}
\bibinfo{author}{{Poutanen}, J.}
\newblock \bibinfo{title}{{Relativistic rotating vector model for X-ray
  millisecond pulsars}}.
\newblock \emph{\bibinfo{journal}{\aap}} \textbf{\bibinfo{volume}{641}},
  \bibinfo{pages}{A166} (\bibinfo{year}{2020}).

\bibitem{Buchner14}
\bibinfo{author}{{Buchner}, J.} \emph{et~al.}
\newblock \bibinfo{title}{{X-ray spectral modelling of the AGN obscuring region
  in the CDFS: Bayesian model selection and catalogue}}.
\newblock \emph{\bibinfo{journal}{\aap}} \textbf{\bibinfo{volume}{564}},
  \bibinfo{pages}{A125} (\bibinfo{year}{2014}).

\bibitem{Brown1978}
\bibinfo{author}{{Brown}, J.~C.}, \bibinfo{author}{{McLean}, I.~S.} \&
  \bibinfo{author}{{Emslie}, A.~G.}
\newblock \bibinfo{title}{{Polarisation by Thomson scattering in optically thin
  stellar envelopes. II. Binary and multiple star envelopes and the
  determination of binary inclinations.}}
\newblock \emph{\bibinfo{journal}{\aap}} \textbf{\bibinfo{volume}{68}},
  \bibinfo{pages}{415--427} (\bibinfo{year}{1978}).

\bibitem{Kravtsov20}
\bibinfo{author}{{Kravtsov}, V.} \emph{et~al.}
\newblock \bibinfo{title}{{Orbital variability of the optical linear
  polarization of the {\ensuremath{\gamma}}-ray binary LS I +61{\textdegree}
  303 and new constraints on the orbital parameters}}.
\newblock \emph{\bibinfo{journal}{\aap}} \textbf{\bibinfo{volume}{643}},
  \bibinfo{pages}{A170} (\bibinfo{year}{2020}).

\bibitem{1997MNRAS.288...43R}
\bibinfo{author}{{Reynolds}, A.~P.} \emph{et~al.}
\newblock \bibinfo{title}{{A new mass estimate for Hercules X-1}}.
\newblock \emph{\bibinfo{journal}{\mnras}} \textbf{\bibinfo{volume}{288}},
  \bibinfo{pages}{43--52} (\bibinfo{year}{1997}).

\bibitem{leahy14}
\bibinfo{author}{{Leahy}, D.~A.} \& \bibinfo{author}{{Abdallah}, M.~H.}
\newblock \bibinfo{title}{{HZ Her: Stellar Radius from X-Ray Eclipse
  Observations, Evolutionary State, and a New Distance}}.
\newblock \emph{\bibinfo{journal}{\apj}} \textbf{\bibinfo{volume}{793}},
  \bibinfo{pages}{79} (\bibinfo{year}{2014}).

\bibitem{2016ApJ...831..194W}
\bibinfo{author}{{Wolff}, M.~T.} \emph{et~al.}
\newblock \bibinfo{title}{{The NuSTAR X-Ray Spectrum of Hercules X-1: A
  Radiation-dominated Radiative Shock}}.
\newblock \emph{\bibinfo{journal}{\apj}} \textbf{\bibinfo{volume}{831}},
  \bibinfo{pages}{194} (\bibinfo{year}{2016}).

\bibitem{2015MNRAS.447.1847M}
\bibinfo{author}{{Mushtukov}, A.~A.}, \bibinfo{author}{{Suleimanov}, V.~F.},
  \bibinfo{author}{{Tsygankov}, S.~S.} \& \bibinfo{author}{{Poutanen}, J.}
\newblock \bibinfo{title}{{The critical accretion luminosity for magnetized
  neutron stars}}.
\newblock \emph{\bibinfo{journal}{\mnras}} \textbf{\bibinfo{volume}{447}},
  \bibinfo{pages}{1847--1856} (\bibinfo{year}{2015}).

\bibitem{2007A&A...465L..25S}
\bibinfo{author}{{Staubert}, R.} \emph{et~al.}
\newblock \bibinfo{title}{{Discovery of a flux-related change of the cyclotron
  line energy in Hercules X-1}}.
\newblock \emph{\bibinfo{journal}{\aap}} \textbf{\bibinfo{volume}{465}},
  \bibinfo{pages}{L25--L28} (\bibinfo{year}{2007}).

\bibitem{1969SvA....13..175Z}
\bibinfo{author}{{Zel'dovich}, Y.~B.} \& \bibinfo{author}{{Shakura}, N.~I.}
\newblock \bibinfo{title}{{X-Ray Emission Accompanying the Accretion of Gas by
  a Neutron Star}}.
\newblock \emph{\bibinfo{journal}{\sovast}} \textbf{\bibinfo{volume}{13}},
  \bibinfo{pages}{175} (\bibinfo{year}{1969}).

\bibitem{2018A&A...619A.114S}
\bibinfo{author}{{Suleimanov}, V.~F.}, \bibinfo{author}{{Poutanen}, J.} \&
  \bibinfo{author}{{Werner}, K.}
\newblock \bibinfo{title}{{Accretion heated atmospheres of X-ray bursting
  neutron stars}}.
\newblock \emph{\bibinfo{journal}{\aap}} \textbf{\bibinfo{volume}{619}},
  \bibinfo{pages}{A114} (\bibinfo{year}{2018}).

\bibitem{2019MNRAS.483..599G}
\bibinfo{author}{{Gonz{\'a}lez-Caniulef}, D.}, \bibinfo{author}{{Zane}, S.},
  \bibinfo{author}{{Turolla}, R.} \& \bibinfo{author}{{Wu}, K.}
\newblock \bibinfo{title}{{Atmosphere of strongly magnetized neutron stars
  heated by particle bombardment}}.
\newblock \emph{\bibinfo{journal}{\mnras}} \textbf{\bibinfo{volume}{483}},
  \bibinfo{pages}{599--613} (\bibinfo{year}{2019}).

\bibitem{2021MNRAS.503.5193M}
\bibinfo{author}{{Mushtukov}, A.~A.}, \bibinfo{author}{{Suleimanov}, V.~F.},
  \bibinfo{author}{{Tsygankov}, S.~S.} \& \bibinfo{author}{{Portegies Zwart},
  S.}
\newblock \bibinfo{title}{{Spectrum formation in X-ray pulsars at very low mass
  accretion rate: Monte Carlo approach}}.
\newblock \emph{\bibinfo{journal}{\mnras}} \textbf{\bibinfo{volume}{503}},
  \bibinfo{pages}{5193--5203} (\bibinfo{year}{2021}).

\bibitem{2006MNRAS.373.1495V}
\bibinfo{author}{{van Adelsberg}, M.} \& \bibinfo{author}{{Lai}, D.}
\newblock \bibinfo{title}{{Atmosphere models of magnetized neutron stars: QED
  effects, radiation spectra and polarization signals}}.
\newblock \emph{\bibinfo{journal}{\mnras}} \textbf{\bibinfo{volume}{373}},
  \bibinfo{pages}{1495--1522} (\bibinfo{year}{2006}).

\bibitem{2009A&A...500..891S}
\bibinfo{author}{{Suleimanov}, V.}, \bibinfo{author}{{Potekhin}, A.~Y.} \&
  \bibinfo{author}{{Werner}, K.}
\newblock \bibinfo{title}{{Models of magnetized neutron star atmospheres: thin
  atmospheres and partially ionized hydrogen atmospheres with vacuum
  polarization}}.
\newblock \emph{\bibinfo{journal}{\aap}} \textbf{\bibinfo{volume}{500}},
  \bibinfo{pages}{891--899} (\bibinfo{year}{2009}).

\bibitem{2000ApJ...529.1011P}
\bibinfo{author}{{Pavlov}, G.~G.} \& \bibinfo{author}{{Zavlin}, V.~E.}
\newblock \bibinfo{title}{{Polarization of Thermal X-Rays from Isolated Neutron
  Stars}}.
\newblock \emph{\bibinfo{journal}{\apj}} \textbf{\bibinfo{volume}{529}},
  \bibinfo{pages}{1011--1018} (\bibinfo{year}{2000}).

\end{thebibliography}
\end{document}